\documentclass[prd,twocolumn,showpacs,superscriptaddress,nofootinbib,floatfix,showkeys,10pt]{revtex4-2}

\usepackage{tikz}
\usepackage{tikz-feynman} 
\usepackage{adjustbox}
\usepackage{graphicx}
\usepackage{amsmath}
\usepackage{bm}
\usepackage{tabularx}
\usepackage{yhmath}
\usepackage{mathtools}
\usepackage{float}
\usepackage{wasysym}
\usepackage{color,xcolor}
\usepackage{subfigure}
\usepackage{cases}
\usepackage{subfigure}
\usepackage{times}
\usepackage{dcolumn,booktabs,bm}
\usepackage{slashed}
\usepackage{amsfonts,amssymb,stmaryrd,latexsym}
\usepackage{mathrsfs}
\usepackage{textcomp}
\usepackage{multirow}
\usepackage{cancel}
\usepackage{array}
\usepackage{enumerate,enumitem}
\usepackage{orcidlink}
\usepackage{dashrule}
\usepackage{multirow}
\usepackage{times}
\usepackage{mdframed}
\usepackage{hyperref}
\usepackage[titletoc]{appendix}

\definecolor{mycolor}{RGB}{45,48,146}
\hypersetup{colorlinks=true,citecolor=mycolor,urlcolor=mycolor,linkcolor=mycolor}

\allowdisplaybreaks[4]
\newcounter{cnt}
\makeatletter
\let\oldhypertarget\hypertarget
\renewcommand{\hypertarget}[2]{%
	\oldhypertarget{#1}{#2}%
	\protected@write\@mainaux{}{%
		\string\expandafter\string\gdef
		\string\csname\string\detokenize{#1}\string\endcsname{#2}%
	}%
}
\newcommand{\myhyperlink}[1]{%
	\hyperlink{#1}{\csname #1\endcsname}%
}
\makeatother

\begin{document}

\title{Strong decays of the hidden-charm molecular pentaquarks}
	
\author{Jin-Cheng Deng}\thanks{These authors equally contribute to this work.}
\affiliation{College of Physics Science and Technology, Hebei University, Baoding 071002, China}
	
\author{Yong Ru}\thanks{These authors equally contribute to this work.}
\affiliation{College of Physics Science and Technology, Hebei University, Baoding 071002, China}
	
\author{Xin-Yue Wan}
\affiliation{College of Physics Science and Technology, Hebei University, Baoding 071002, China}

\author{Tai-Fu Feng}\email{fengtf@hbu.edu.cn}
\affiliation{College of Physics Science and Technology, Hebei University, Baoding 071002, China}
\affiliation{Hebei Key Laboratory of High-precision Computation and Application of Quantum Field Theory, Baoding, 071002, China}
\affiliation{Hebei Research Center of the Basic Discipline for Computational Physics, Baoding, 071002, China}
	
\author{Bo Wang\,\orcidlink{0000-0003-0985-2958}}\email{wangbo@hbu.edu.cn}
\affiliation{College of Physics Science and Technology, Hebei University, Baoding 071002, China}
\affiliation{Hebei Key Laboratory of High-precision Computation and Application of Quantum Field Theory, Baoding, 071002, China}
\affiliation{Hebei Research Center of the Basic Discipline for Computational Physics, Baoding, 071002, China}
	
\begin{abstract}
We investigate the strong decays of the recently observed hidden-charm pentaquarks \(P_{\psi}^N(4312)\), \(P_{\psi}^N(4440)\), and \(P_{\psi}^N(4457)\), as well as \(P_{\psi s}^\Lambda(4338)\) and \(P_{\psi s}^\Lambda(4459)\), within the molecular framework using the effective Lagrangian approach. We construct the effective Lagrangians describing the S-wave couplings between these pentaquarks and their constituent hadrons, namely \(\Sigma_c\bar{D}^{(*)}\) and \(\Xi_c\bar{D}^{(*)}\), and determine the coupling constants via the residues of the scattering \(T\)-matrix at the bound-state poles. Our results show that the decay widths are sensitive to the cutoff parameters in the form factors, whereas the branching fractions exhibit only weak dependence. Using \(P_{\psi}^N(4312)\) to calibrate the cutoff range, we further explore the spin assignments of \(P_{\psi}^N(4440)\) and \(P_{\psi}^N(4457)\). Our calculations favor the assignment where the lower-mass state \(P_{\psi}^N(4440)\) carries higher spin \(J = 3/2\) and the higher-mass state \(P_{\psi}^N(4457)\) carries lower spin \(J = 1/2\). In addition, the experimental widths of \(P_{\psi s}^\Lambda(4338)\) and \(P_{\psi s}^\Lambda(4459)\) can both be well reproduced under the molecular interpretation. We look forward to future experimental analyses with more accumulated data to clarify whether the lineshape of \(P_{\psi s}^\Lambda(4459)\) contains contributions from both spin-\(1/2\) and spin-\(3/2\) states.
\end{abstract}
	
\maketitle
	
\section{Introduction}\label{sec:intro}

A significant impetus for the remarkable progress in hadron physics over the past two decades has been the experimental discovery of a plethora of near-threshold exotic states~\cite{ParticleDataGroup:2024cfk}. These states defy a simple interpretation within the conventional quark model, where mesons are described as quark-antiquark ($q\bar{q}$) pairs and baryons as three-quark ($qqq$) systems. Experimental evidence strongly suggests that their internal structure comprises a higher valence quark content, such as tetraquark ($qq\bar{q}\bar{q}$) or pentaquark ($qqqq\bar{q}$) configurations. While the emergence of exotic states challenges the traditional quark model, it also motivates a deeper and broader perspective on low-energy strong interactions. This has led to the compelling picture that color-neutral hadrons can bind together through residual strong forces to form a new hierarchical structure of matter—the hadronic molecular state~\cite{Chen:2016qju,Guo:2017jvc,Liu:2019zoy,Lebed:2016hpi,Esposito:2016noz,Brambilla:2019esw,Yang:2020atz,Chen:2021ftn,Chen:2022asf,Meng:2022ozq,Liu:2024uxn,Wang:2025sic}. The deuteron, a bound state of a proton and a neutron, serves as a classic archetype for such a system.  Moreover, extensive theoretical studies have shown that heavy-flavor hadrons can form molecular states via attractive interactions in specific channels, a scenario that aligns well with the molecular candidates reported in experiments in recent years.

In 2019, the LHCb Collaboration reported the discovery of three narrow pentaquark states, namely $P_{\psi}^N(4312)^+$, $P_{\psi}^N(4440)^+$, and $P_{\psi}^N(4457)^+$, in the $J/\psi p$ invariant mass spectrum of the $\Lambda_b^0 \to J/\psi p K^-$ decay~\cite{LHCb:2019kea}. This observation resolved the broad $P_{\psi}^N(4450)^+$ structure, first observed in 2015~\cite{LHCb:2015yax}, into two distinct narrower states: $P_{\psi}^N(4440)^+$ and $P_{\psi}^N(4457)^+$. Their experimentally measured masses and widths $[m,\Gamma]$ are, respectively:
\begin{eqnarray}
P_{\psi}^N(4312)^+:&& \left[4311.9 \pm 0.7_{-0.6}^{+6.8},~ 9.8 \pm 2.7_{-4.5}^{+3.7}\right] ~\mathrm{MeV},\nonumber\\
P_{\psi}^N(4440)^+:&& \left[4440.3 \pm 1.3_{-4.7}^{+4.1},~ 20.6 \pm 4.9_{-10.1}^{+8.7}\right] ~\mathrm{MeV},\nonumber\\
P_{\psi}^N(4457)^+:&& \left[4457.3 \pm 0.6_{-1.7}^{+4.1},~ 6.4 \pm 2.0_{-1.9}^{+5.7}\right]~\mathrm{MeV}.\nonumber
\end{eqnarray}

The discovery of these pentaquark states has ignited extensive investigations into their properties~\cite{Meng:2019ilv,Chen:2019bip,Liu:2019tjn,Chen:2019asm,Xiao:2019aya,He:2019ify,Xiao:2019mvs,Guo:2019fdo,Guo:2019kdc,Weng:2019ynv,Burns:2019iih,Wang:2019ato,Du:2019pij,Wang:2019got,Wang:2019spc,Xu:2020gjl,Yalikun:2021bfm, Xing:2022ijm,Zhang:2023czx,Cheng:2019obk}, such as their mass spectrum, decay patterns, and production mechanisms. Currently, three main interpretations have been proposed to explain their nature~\cite{Chen:2016qju,Guo:2017jvc,Liu:2019zoy,Lebed:2016hpi,Esposito:2016noz,Brambilla:2019esw,Yang:2020atz,Chen:2021ftn,Chen:2022asf,Meng:2022ozq,Liu:2024uxn}: the hadronic molecular state, the compact multiquark (pentaquark) state, and kinematic effects. Among these, the molecular picture has emerged as the most prevalent, primarily because the masses of these three states are located just a few to several MeV below the $\Sigma_c\bar{D}$ and $\Sigma_c\bar{D}^\ast$ thresholds, respectively. This proximity is in good agreement with the expectations of the molecular hypothesis. However, the current experimental data lack sufficient statistics to definitively determine their spin-parity ($J^P$) quantum numbers~\cite{LHCb:2019kea}. Within the molecular framework, an S-wave interaction between the $\Sigma_c\bar{D}$ constituents can form a state with total angular momentum $J=1/2$, while the $\Sigma_c\bar{D}^\ast$ system can form states with $J=1/2$ and $J=3/2$. Consequently, assuming negative parity from the S-wave nature, the $J^P$ of $P_{\psi}^N(4312)^+$ is theoretically assigned as $1/2^-$. For $P_{\psi}^N(4440)^+$ and $P_{\psi}^N(4457)^+$, it remains ambiguous which corresponds to the $1/2^-$ state and which to the $3/2^-$ state. A primary motivation of the present work is to investigate the quantum number assignments of the latter two states by studying their strong decay behaviors.

The discovery of the $P_{\psi}^N$ states has further spurred theoretical and experimental searches for their strange partners—pentaquarks containing a strange quark, anticipated in the energy region of the $\Xi_c^{(\prime,\ast)}\bar{D}^{(\ast)}$ systems. For instance, studies in Refs.~\cite{Feijoo:2015kts,Chen:2015sxa,Lu:2016roh,Chen:2016ryt,Xiao:2019gjd,Wang:2019nvm} predicted the existence of molecular states in the isospin $I=0$, $\Xi_c^{(\prime,\ast)}\bar{D}^{(\ast)}$ systems. In 2021, the LHCb Collaboration found evidence for a new state, $P_{\psi s}^\Lambda(4459)$, in the $J/\psi\Lambda$ invariant mass spectrum from the decay $\Xi_b^-\to J/\psi\Lambda K^-$~\cite{LHCb:2020jpq}. Its measured mass and width are:
\begin{eqnarray}
P_{\psi s}^\Lambda(4459):&& \left[4458.8 \pm 2.9_{-1.1}^{+4.7},~ 17.3 \pm 6.5_{-5.7}^{+8.0}\right] ~\mathrm{MeV}.\nonumber
\end{eqnarray}
This result is in remarkable agreement with our predictions from chiral effective field theory in Ref.~\cite{Wang:2019nvm}. The mass of the $P_{\psi s}^\Lambda(4459)$ lies in close proximity to the $\Xi_c\bar{D}^\ast$ threshold, making it an good candidate for a $\Xi_c\bar{D}^\ast$ molecular state. An S-wave interaction in the $\Xi_c\bar{D}^\ast$ system can produce states with total angular momentum $J=1/2$ and $J=3/2$. The LHCb collaboration performed a fit to the $J/\psi\Lambda$ invariant mass spectrum under a double-peak hypothesis~\cite{Wang:2019nvm}, but the current statistical significance is insufficient to determine whether the $P_{\psi s}^\Lambda(4459)$ peak indeed contains two underlying sub-structures.

Subsequently, the LHCb Collaboration in 2023 discovered the $P_{\psi s}^\Lambda(4338)$ state in the $J/\psi\Lambda$ invariant mass spectrum of the $B^-\to J/\psi\Lambda\bar{p}$ decay~\cite{LHCb:2022ogu}, with its mass and width measured to be:
\begin{eqnarray}
P_{\psi s}^\Lambda(4338):&& \left[4338.2 \pm 0.7 \pm 0.4,~ 7.0 \pm 1.2 \pm 1.3\right] ~\mathrm{MeV}.\nonumber
\end{eqnarray}
The experimental analysis favors a $J^P$ quantum number of $1/2^-$. We noted that the mass of the $P_{\psi s}^\Lambda(4338)$ is exceptionally close to the $\Xi_c\bar{D}$ threshold. This behavior is in stark contrast to the predictions of most theoretical models. We demonstrated in Refs.~\cite{Wang:2023hpp,Wang:2023eng} that by employing an extended heavy quark symmetry—wherein the strange quark within the $\Xi_c$ baryon is also treated as a heavy quark—one can establish a connection between the $P_{\psi s}^\Lambda(4338)$ and the recently discovered $T_{cc}(3875)^+$ within the molecular framework. This approach allows us to naturally explain the near-threshold nature of the $P_{\psi s}^\Lambda(4338)$.

The discoveries of $P_{\psi s}^\Lambda(4459)$ and $P_{\psi s}^\Lambda(4338)$ have triggered a flurry of research activities focusing on their production mechanisms~\cite{Wu:2021caw,Cheng:2021gca,Paryev:2023icm}, mass spectra~\cite{Chen:2020uif,Peng:2020hql,Wang:2020eep,Liu:2020hcv,Dong:2021juy,Wang:2021hql,Zhu:2021lhd,Xiao:2021rgp,Giron:2021sla,Hu:2021nvs,Chen:2021cfl,Chen:2021spf,Chen:2022onm}, decay properties~\cite{Chen:2021tip,Lu:2021irg,Yang:2021pio,Azizi:2021pbh}, experimental lineshapes~\cite{Du:2021bgb,Meng:2022wgl,Burns:2022uha,Nakamura:2022gtu}, as well as the magnetic moments~\cite{Ozdem:2021ugy,Li:2021ryu,Gao:2021hmv,Wang:2022tib,Ozdem:2022kei}.

In addition to the mass spectrum, the decay characteristics of pentaquarks are intimately linked to their internal structure. Consequently, a systematic investigation of the strong decay behaviors of the hidden-charm pentaquarks observed by LHCb will be instrumental in discerning their underlying nature, particularly when confronted with forthcoming experimental data. In the present work, we perform a systematic calculation of the strong decays for both the $P_{\psi}^N$ and $P_{\psi s}^\Lambda$ states within the hadronic molecular framework. Our primary objective is to leverage the crucial information provided by decay branching fractions to discriminate between the $J^P$ assignments of $P_{\psi}^N(4440)^+$ and $P_{\psi}^N(4457)^+$. Furthermore, we aim to explore the distinct decay patterns arising from the $\Xi_c\bar{D}^*$ system in its $J=1/2$ and $J=3/2$ configurations.

The remainder of this paper is structured as follows. In Sec.~\ref{sec:effl}, we delineate our theoretical framework, which encompasses the effective Lagrangians employed, the approach for determining the coupling constants between the pentaquarks and their constituents, and the formulation of the decay amplitudes. In Sec.~\ref{sec:nums}, we present and discuss our numerical results. Finally, a summary of our work is provided in Sec.~\ref{sec:sum}.

\section{Effective Lagrangians, coupling constants, and decay amplitudes}\label{sec:effl}
	
\subsection{Effective Lagrangians}\label{sec:efflags}

\begin{figure*}[htbp]
\begin{centering}
    \scalebox{1.0}{\includegraphics[width=\linewidth]{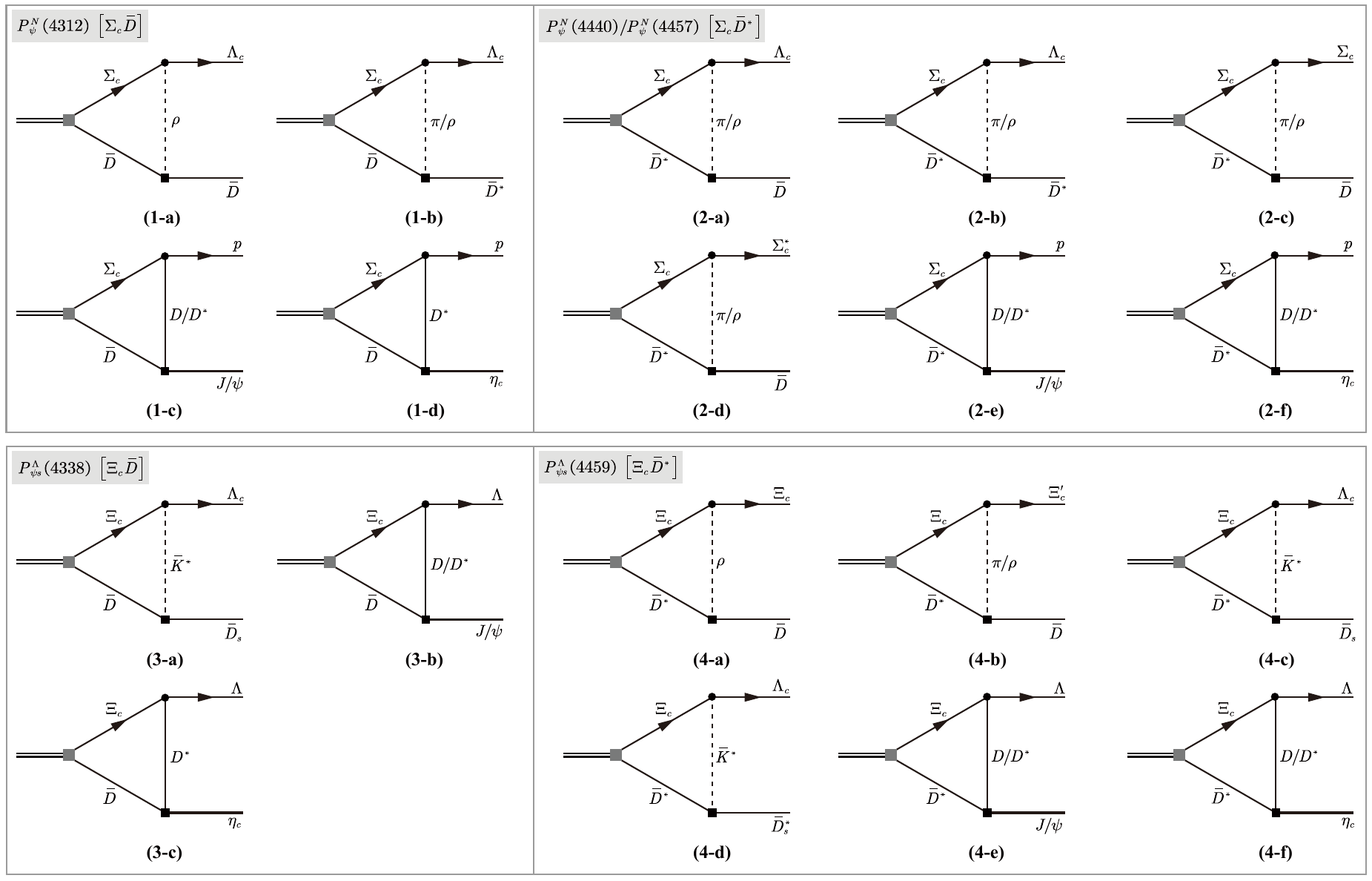}}
    \caption{Triangle loop diagrams for the strong decays of the $P_{\psi}^N$ and $P_{\psi s}^\Lambda$ states. The double lines denote the corresponding pentaquarks.\label{fig:decays}}
\end{centering}
\end{figure*}

In Fig.~\ref{fig:decays}, we present the Feynman diagrams for the strong decays of the $P_{\psi}^N$ and $P_{\psi s}^\Lambda$ states via meson exchange. From these, it can be seen that we require the following types of Lagrangians.

\textbf{(i)} We first need to construct the effective Lagrangians that describe the S-wave interactions between the molecular hidden-charm pentaquarks and their constituent particles, which can be written as follows:
\begin{eqnarray}
\mathcal{L}_{P_\psi^N\Sigma_c\bar{D}}&=&g_{\psi}\bar{P}_\psi^N \Sigma_c(i\tau_2)\tilde{D}+\mathrm{H.c.},\label{eq:5} \\ 
\mathcal{L}_{P_\psi^N\Sigma_c\bar{D}^\ast}&=&
\tilde{g}_{\psi}\bar{P}_\psi^N \gamma^\mu\gamma_5\Sigma_c(i\tau_2)\tilde{D}_\mu^\ast \nonumber\\ 
&& +\tilde{g}_{\psi}^\prime\bar{P}_\psi^{N\mu} \Sigma_c(i\tau_2)\tilde{D}_\mu^\ast+\mathrm{H.c.},\label{eq:6} \\
\mathcal{L}_{P_{\psi s}^\Lambda\Xi_c\bar{D}}&=&g_{\psi s}\bar{P}_{\psi s}^\Lambda \Xi_c(i\tau_2)\tilde{D}+\mathrm{H.c.}, \label{eq:7}\\ 
\mathcal{L}_{P_{\psi s}^\Lambda\Xi_c\bar{D}^\ast}&=&
\tilde{g}_{\psi s}\bar{P}_{\psi s}^\Lambda \gamma^\mu\gamma_5\Xi_c(i\tau_2)\tilde{D}_\mu^\ast \nonumber\\ 
&& +\tilde{g}_{\psi s}^\prime\bar{P}_{\psi s}^{\Lambda\mu} \Xi_c(i\tau_2)\tilde{D}_\mu^\ast+\mathrm{H.c.}, \label{eq:8}
\end{eqnarray}
where the notations $P_{\psi}^N/P_{\psi s}^\Lambda$ and $P_{\psi}^{N\mu}/P_{\psi s}^{\Lambda\mu}$ denote the $J=1/2$ and $J=3/2$ spinors of the corresponding pentaquarks, respectively. $\tau_2$ is the second Pauli matrix in isospin space, and $\Sigma_c$ denotes the isospin triplet
\begin{eqnarray}\label{eq:sigc}
\Sigma_c&=&\left[\begin{array}{cc}
\Sigma_c^{++} & \frac{\Sigma_c^{+}}{\sqrt{2}}\\
\frac{\Sigma_c^{+}}{\sqrt{2}} & \Sigma_c^0
\end{array}\right].
\end{eqnarray}
The isospin-$1/2$ fields are respectively given as
\begin{eqnarray}
\bar{P}_\psi^{N(\mu)}&=&(\bar{P}_\psi^{N(\mu)+},\bar{P}_\psi^{N(\mu)0}),\nonumber\\
\tilde{D}^{(\ast)}_{(\mu)}&=&(\bar{D}^{(\ast)0},D^{(\ast)-})^T_{(\mu)},\qquad\Xi_c=(\Xi_c^+,\Xi_c^0).
\end{eqnarray}
The coupling constants appearing in Eqs.~\eqref{eq:5}-\eqref{eq:8} will be determined from the residues of the scattering $T$-matrix in Sec.~\ref{sec:coupg0}.

\textbf{(ii)} The effective Lagrangians describing the baryon-baryon-meson coupling vertices are given by~\cite{Liu:2011xc}:
\begin{eqnarray}
\mathcal{L}_{B_{\bar{3}}} &=& i\beta_B \mathrm{tr}[\bar{B}_{\bar{3}}v^\mu(\Gamma_\mu-V_\mu)B_{\bar{3}}],\label{eq:12} \\
\mathcal{L}_S&=&\frac{3}{2}g_1(iv_\kappa)\epsilon^{\mu\nu\lambda\kappa}\mathrm{tr}[\bar{S}_\mu u_\nu S_\lambda]\nonumber\\
&&+i\beta_S \mathrm{tr}[\bar{S}_\mu v_\alpha(\Gamma^\alpha-V^\alpha)S^\mu]\nonumber\\
&&+\lambda_S \mathrm{tr}[\bar{S}_\mu F^{\mu\nu}S_\nu],\label{eq:13}\\	
\mathcal{L}_{\mathrm{int}}&=&i\lambda_I\epsilon^{\mu\nu\lambda\kappa}v_\mu \mathrm{tr}[\bar{S}_\nu F_{\lambda\kappa}B_{\bar{3}}]\nonumber\\ 
&&+g_4 \mathrm{tr}[\bar{S}^\mu u_\mu B_{\bar{3}}]+\mathrm{H.c.},\label{eq:14}
\end{eqnarray}
where $\mathrm{tr}[\bullet]$ denotes the trace in flavor space, and
\begin{eqnarray}
S_\mu=B_{6\mu}^\ast-\frac{1}{\sqrt{3}}(\gamma_\mu+v_\mu)\gamma_5 B_6.
\end{eqnarray}
The $B_{\bar{3}}$, $B_6$ and $B^*_6$ denote the spin-$\frac{1}{2}$ antitriplet, spin-$\frac{1}{2}$ and spin-$\frac{3}{2}$ sextet of charmed baryons, respectively,
\begin{eqnarray}
B_{\bar{3}}&=&\left[\begin{array}{ccc}
0 & \Lambda_c^+ & \Xi_c^+ \\-\Lambda_c^+ & 0 & \Xi_c^0 \\ -\Xi_c^+ & -\Xi_c^0 & 0 
\end{array}\right],	\\ B_6&=&\left[\begin{array}{ccc}
\Sigma_c^{ + +} & \frac{1}{\sqrt{2}}\Sigma_c^+ & \frac{1}{\sqrt{2}}\Xi_c^{\prime+} \\\frac{1}{\sqrt{2}}\Sigma_c^+ & \Sigma_c^0 & \frac{1}{\sqrt{2}}\Xi_c^{\prime 0} \\ \frac{1}{\sqrt{2}}\Xi_c^{\prime+} &\frac{1}{\sqrt{2}}\Xi_c^{\prime0} &\Omega_c^0 
\end{array}\right],	\\ 	B_6^\ast&=&\left[\begin{array}{ccc}
\Sigma_c^{\ast + +} & \frac{1}{\sqrt{2}}\Sigma_c^{\ast+} & \frac{1}{\sqrt{2}}\Xi_c^{\ast+} \\\frac{1}{\sqrt{2}}\Sigma_c^{\ast+} & \Sigma_c^{\ast0} & \frac{1}{\sqrt{2}}\Xi_c^{\ast0} \\ \frac{1}{\sqrt{2}}\Xi_c^{\ast+} &\frac{1}{\sqrt{2}}\Xi_c^{\ast0} &\Omega_c^{\ast0} 
\end{array}\right].
\end{eqnarray}
The axial-vector current $u_\mu$ and chiral connection $\Gamma_\mu$ are defined as
\begin{eqnarray}
u_\mu=\frac{i}{2}\left\{\xi^\dagger,\partial_\mu\xi\right\},\qquad \Gamma_\mu=\frac{1}{2}\left[\xi^\dagger,\partial_\mu\xi\right],
\end{eqnarray}
with 
\begin{eqnarray}
\xi^2&=&U=\exp\left(\frac{i\varphi}{f_\pi}\right),\nonumber\\ \varphi&=&\sqrt{2}\left[\begin{array}{ccc}
\frac{\pi^0}{\sqrt{2}}+\frac{\eta}{\sqrt{6}} & \pi^+  & K^+  \\
\pi^- & -\frac{\pi^0}{\sqrt{2}}+\frac{\eta}{\sqrt{6}}  & K^0 \\ K^- & \bar{K}^0 &-\frac{2}{\sqrt{6}}\eta
\end{array}\right],
\end{eqnarray}
and the pion decay constant $f_\pi=92.4$ MeV. The antisymmetric tensors $\sigma^{\mu\nu}$ and $F_{\mu\nu}$ are given by
\begin{eqnarray}
\sigma^{\mu\nu}&=&\frac{i}{2}\left[\gamma^\mu,\gamma^\nu\right],\\
F_{\mu\nu}&=&\partial_{\mu}\rho_{\nu}-\partial_{\nu}\rho_{\mu}+\left[\rho_{\mu},\rho_{\nu}\right],
\end{eqnarray}
with
\begin{eqnarray}
\rho_{\mu}=i\frac{g_{V}}{\sqrt{2}}V_\mu,\quad V_\mu&=&\left[\begin{array}{ccc}
\frac{\omega+\rho^0}{\sqrt{2}} & \rho^+ &K^{\ast +}\\
\rho^- & \frac{\omega-\rho^0}{\sqrt{2}} &K^{\ast 0} \\ K^{\ast -}& \bar{K}^{\ast 0} & \phi
\end{array}\right]_\mu.
\end{eqnarray}
The constants $g_V=m_\rho/(\sqrt{2}f_\pi)=5.9$, $\beta_B g_V=-6.0$,  $\beta_S g_V=12.0$,  $\lambda_{S}g_V=19.2$ GeV$^{-1}$ , $\lambda_{I}g_V=-(\lambda_{S}g_V)/\sqrt{8}$ , and $g_1=1.0$ is estimated from the chiral quark model~\cite{Liu:2011xc}, while $g_4=0.99$ is extracted from the partial decay width of the $\Sigma_c^\ast\to \Lambda_c \pi$ process~\cite{ParticleDataGroup:2024cfk,Liu:2011xc}.

The effective Lagrangian describing the $\Sigma_c D^{(\ast)}N$~\cite{Lu:2016nnt} and $\Xi_c D^{(\ast)}\Lambda$~\cite{Clymton:2025zer} couplings take the following forms, respectively:
\begin{eqnarray}
\mathcal{L}_{\Sigma_c D^{(\ast)}N}&=&ig_{\Sigma_c D N} \bar{N} \gamma^5 \Sigma_c (i\tau_2) D^\dagger \nonumber\\
&&+g_{\Sigma_c D^\ast N} \bar{N}\gamma^\mu \Sigma_c (i\tau_2) D^{\ast\dagger}_\mu +\mathrm{H.c.},\label{eq:31}\\
\mathcal{L}_{B_8 B_{\bar{3}} D^{(\ast)}}&=&ig_{B_8 B_{\bar{3}} D}\mathrm{tr}[\bar{B}_8\gamma^5 (D^\dagger B_{\bar{3}})_8] \nonumber\\	
&&+g_{B_8 B_{\bar{3}} D^\ast}\mathrm{tr}[\bar{B}_8\gamma^\mu (D_\mu^{\ast\dagger} B_{\bar{3}})_8]+\mathrm{H.c.},\label{eq:34}
\end{eqnarray}
where the $N =(p, n)^T$, $D^{(\ast)}=(D^{(\ast)0},D^{(\ast)+})$, and the matrix $\Sigma_c$ is given in Eq.~\eqref{eq:sigc}. The baryon octet $B_8$ is given by
\begin{eqnarray}
B_8&=&\left[\begin{array}{ccc}
\frac{\Sigma^0}{\sqrt{2}}+\frac{\Lambda}{\sqrt{6}} & \Sigma^+  &p  \\
\Sigma^- & -\frac{\Sigma^0}{\sqrt{2}}+\frac{\Lambda}{\sqrt{6}}  & n \\ \Xi^- & \Xi^0 &-\frac{2}{\sqrt{6}}\Lambda
\end{array}\right], 
\end{eqnarray}
and
\begin{widetext}
\begin{eqnarray}
(D^\dagger B_{\bar{3}})_8&=&\left[\begin{array}{ccc}
\frac{2}{3}D^{0\dagger}\Xi_c^0+\frac{1}{3}D^{+\dagger}\Xi_c^+-\frac{1}{3}D_s^{+\dagger}\Lambda_c^+& -D^{0\dagger}\Xi_c^+  &D^{0\dagger}\Lambda_c^+  \\
D^{+\dagger}\Xi_c^0 & 	-\frac{2}{3}D^{+\dagger}\Xi_c^+-\frac{1}{3}D^{0\dagger}\Xi_c^0-\frac{1}{3}D_s^{+\dagger}\Lambda_c^+  &D^{+\dagger}\Lambda_c^+ \\ 
D_s^{+\dagger}\Xi_c^0 & -D_s^{+\dagger}\Xi^+&\frac{2}{3}D_s^{+\dagger}\Lambda_c^+ -\frac{1}{3}D^{0\dagger}\Xi_c^0+\frac{1}{3}D^{+\dagger}\Xi_c^+
\end{array}\right].\nonumber\\
\end{eqnarray}
\end{widetext}
The treatment of $(D^{\ast\dagger} B_{\bar{3}})_8$ follows an analogous structure. The coupling constants $g_{\Sigma_c D N}=2.69$ and $g_{\Sigma_c D^\ast N}=4.24$ are derived from the SU(4) invariant Lagrangians~\cite{Dong:2010xv,Liu:2001ce,Okubo:1975sc} and SU(3) flavor symmetry~\cite{Doring:2010ap,deSwart:1963pdg}, and $g_{B_8 B_{\bar{3}} D}=13.5$ and $g_{B_8 B_{\bar{3}} D^\ast}=14.0$~\cite{Shimizu:2017xrg}.

\textbf{(iii)} The effective Lagrangians for the three-meson coupling vertices are given by~\cite{Casalbuoni:1996pg}
\begin{eqnarray}
\mathcal{L}_\pi &=& g_b\langle\bar{\tilde{\mathcal{H}}}\gamma^\mu\gamma^5u_\mu\tilde{\mathcal{H}}\rangle,\label{eq:24} \\
\mathcal{L}_\rho&=& i\beta \langle \bar{\tilde{\mathcal{H}}}v^{\mu}(\Gamma_{\mu}-\rho_{\mu})\tilde{\mathcal{H}} \rangle 
+i\lambda \langle \bar{\tilde{\mathcal{H}}}\sigma^{\mu\nu}F_{\mu\nu}\tilde{\mathcal{H}} \rangle,\label{eq:25}\\
\mathcal{L}_2&=&g_2 \langle \mathcal{J}\bar{\tilde{\mathcal{H}}}\overleftrightarrow{\slashed{\partial}}\bar{\mathcal{H}}\rangle+\mathrm{H.c.},\label{eq:28}
\end{eqnarray}
where the notation $\langle \bullet\rangle$ denotes the trace in spinor space, and
\begin{eqnarray}
\tilde{\mathcal{H}}&=&\left(\tilde{P}_{\mu}^{\ast}\gamma^{\mu}+i\tilde{P}\gamma^{5}\right)\frac{1-\slashed{v}}{2},\quad \bar{\tilde{\mathcal{H}}}=\gamma^0\tilde{\mathcal{H}}^\dagger\gamma^0,\\
\mathcal{J}&=&\frac{1+\slashed{v}}{2}\left(\psi^\mu\gamma_\mu+i\eta_c\gamma_5\right)\frac{1-\slashed{v}}{2}.\label{eq:supJ}
\end{eqnarray}
$\tilde{P}^{(\ast)}=(\bar{D}^{(\ast)0},D^{(\ast)-},\bar{D}_s^{(\ast)-})^T$ represents the heavy meson multiplet. $\psi^\mu$ and $\eta_c$ in Eq.~\eqref{eq:supJ} denote the $J/\psi(1S)$ and $\eta_c(1S)$, respectively. The coupling constant $|g_b|=0.59$ is extracted from the partial decay width of the $D^\ast$ meson, specifically from the process $D^{\ast+}\to D^0\pi^+$~\cite{ParticleDataGroup:2024cfk}. The constants $\beta=0.9$, and $\lambda=0.56$ GeV$^{-1}$~\cite{Casalbuoni:1996pg}. While the $g_2$ in Eq.~\eqref{eq:28} is related to the decay constants of $J/\psi$ with the vector meson dominance model~\cite{Colangelo:2003sa},
\begin{eqnarray}
g_2=\frac{\sqrt{m_\psi}}{2m_D f_\psi},
\end{eqnarray}
where the $f_\psi$ can be determined through the decay width of $J/\psi\to e^+e^-$~\cite{ParticleDataGroup:2024cfk}, which gives $f_\psi=415$ MeV.

\subsection{Determination of the coupling constants in Eqs.~\eqref{eq:5}-\eqref{eq:8}}\label{sec:coupg0}

In this subsection, we determine the coupling constants appearing in Eqs.~\eqref{eq:5}–\eqref{eq:8} by using the residue of the scattering $T$-matrix at the bound state pole. We first consider the elastic scattering process between a meson $A$ and a baryon $B$, assuming the interaction is a sufficiently attractive potential such that $A$ and $B$ can form a bound state. Thus, the nonperturbative interaction between $A$ and $B$ can be described within the framework of relativistic quantum field theory by the following Bethe-Salpeter equation:
\begin{eqnarray}
\mathcal{T} &=& \mathcal{V}+\mathcal{V}\mathcal{G}\mathcal{T},
\end{eqnarray}
where $\mathcal{G}$ is the two-body propagator for the meson-baryon system, given by the following expression:
\begin{eqnarray}\label{eq:relg}
\mathcal{G}&=&\int\frac{q^2dq}{(2\pi)^{2}} \frac{\omega_A+\omega_B}{\omega_A\omega_B}\frac{2m_B}{s-(\omega_A+\omega_B)^2+i\epsilon},
\end{eqnarray}
where $\omega_A=\sqrt{\bm{q}^2+m_A^2}$, $\omega_B=\sqrt{\bm{q}^2+m_B^2}$, with $m_A$ and $m_B$ the masses of the meson $A$ and baryon $B$, respectively. $s$ denotes the square of the center of mass energy of $AB$ system.
When the center of mass energy is near the threshold, the scattering between $A$ and $B$ particles can also be described by the following nonrelativistic Lippmann–Schwinger equation:
\begin{eqnarray}
  t &=& v+vGt,
\end{eqnarray}
in which the $G$ can be obtained through the nonrelativistic reduction of the $\mathcal{G}$ in Eq.~\eqref{eq:relg}, e.g., see Ref.~\cite{Deng:2024pep},
\begin{eqnarray}
  \mathcal{G} &=& \frac{m_B}{\mu\sqrt{s}}G,
\end{eqnarray}
with
\begin{eqnarray}\label{eq:nonrelg}
G=\int^\Lambda\frac{q^2dq}{(2\pi)^2}\frac{1}{\sqrt{s}-m_A -m_B-\frac{\bm{q}^2}{2\mu}+i\epsilon},
\end{eqnarray}
where $\mu$ is the reduced mass of the $AB$ system. We introduce a sharp cutoff $\Lambda$ to regularize the loop integral $G$. For a bound state, i.e., when \(\sqrt{s} < m_1 + m_2\), the expression for \(G\) after integration takes the form:
\begin{eqnarray}
G(\sqrt{s})&=&\frac{\mu}{2\pi^{2}}\left[\kappa\arctan\left(\frac{\Lambda}{\kappa}\right)-\Lambda\right]\nonumber\\
&\simeq&\frac{\mu}{2\pi^{2}}\left(\frac{\pi}{2}\kappa-\Lambda\right),
\end{eqnarray}
where $\kappa=\sqrt{2\mu(m_{A}+m_{B}-\sqrt{s})}$ represents the binding momentum. In deriving the result in the second line of the above equation, we employed the approximation $\kappa \ll \Lambda$. The linear term in $\Lambda$ is absorbed by the effective potential of the $AB$ system during the renormalization procedure. Detailed discussions can be found in Ref.~\cite{Meng:2021jnw}.

To ensure that the nonrelativistic reduction does not alter the position of the pole in the $T$-matrix, the condition $\mathcal{V}\mathcal{G} = v G$ must be satisfied.  Thus, we can deduce that $\mathcal{V} = \frac{\mu\sqrt{s}}{m_B} v$, and further obtain 
\begin{eqnarray}
\mathcal{T} &=& \frac{\mu\sqrt{s}}{m_B}t.
\end{eqnarray}
Now suppose there exists a bound state pole with mass $m_\mathbb{P}$ in the $T$-matrix. Near this pole, the $T$-matrix can be expressed in the following form:
\begin{eqnarray}\label{eq:relt}
\mathcal{T}=\frac{g^2}{s-m_\mathbb{P}^2},
\end{eqnarray}
where $g$ represents the coupling constant between the bound state $\mathbb{P}$ and its constituents $A$ and $B$. Using the L'Hôpital's rule, $g^2$ can be expressed as:
\begin{eqnarray}\label{eq:gft1}
g^2&=&\lim_{s\to m_{\mathbb{P}}^2}(s-m_{\mathbb{P}}^2)\mathcal{T}\nonumber\\
&=&\frac{2\mu m_{\mathbb{P}}^2}{m_B}\lim_{\sqrt{s}\to m_{\mathbb{P}}}\left[\frac{d}{d\sqrt{s}}t^{-1}(\sqrt{s})\right]^{-1}\nonumber\\
&=&\frac{8\pi\kappa m_\mathbb{P}^2}{\mu m_B},
\end{eqnarray}
where we have used the relation $t^{-1}=v^{-1}-G$, with a energy independent effective potential $v$.

\begin{figure}[htbp]
\begin{centering}
    \scalebox{1.0}{\includegraphics[width=0.6\linewidth]{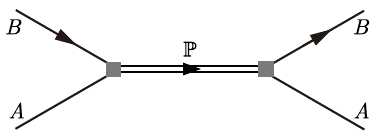}}
    \caption{The elastic scattering process of particles $A$ and $B$, where $\mathbb{P}$ represents the bound state of $A$ and $B$.\label{fig:scatt}}
\end{centering}
\end{figure}

The description in Eq.~\eqref{eq:relt} is equivalent to the near-threshold elastic scattering process illustrated in Fig.~\ref{fig:scatt}. The scattering amplitude for the process shown in Fig.~\ref{fig:scatt} can also be derived from the effective Lagrangians given in Eqs.~\eqref{eq:5}–\eqref{eq:8}. After performing the nonrelativistic reduction, the scattering amplitude is expressed as follows:
\begin{eqnarray}\label{eq:elasm}
i\mathcal{M}_{[B\bar{D}]_J}&=&-i\frac{\mathfrak{g}^{2}}{s-m_\mathbb{P}^2}(2m_\mathbb{P}),\quad J=1/2,\\
i\mathcal{M}_{[B\bar{D}^\ast]_J}&=&
\begin{cases}-i\frac{\mathfrak{g}^{2}}{s-m_\mathbb{P}^2}(6m_\mathbb{P}), &J=1/2, \\
-i\frac{\mathfrak{g}^{2}}{s-m_\mathbb{P}^2}(2m_\mathbb{P}), &J=3/2,
\end{cases}
\end{eqnarray}
where $B$ denotes \(\Sigma_c\) and \(\Xi_c\), and \(m_\mathbb{P}\) represents the mass of the $AB$ bound state with spin \(J\). Once the particles $A$ and $B$ are specified, the coupling constant \(\mathfrak{g}\) corresponds to the relevant coupling constant in the Lagrangians given by Eqs.~\eqref{eq:5}–\eqref{eq:8}.

Using the matching condition
\begin{eqnarray}
\mathcal{M}_{[B\bar{D}^{(\ast)}]_J}&=&-\frac{g^2}{s-m_\mathbb{P}^2},
\end{eqnarray}
one can easily obtain the coupling constants in Eqs.~\eqref{eq:5}–\eqref{eq:8}. We summarize their forms and values in Table~\ref{tab:gvalues}. As can be seen from the numerical results in Table~\ref{tab:gvalues}, the values of these dimensionless coupling constants are all around $1$, which satisfies the requirement of naturalness.

\begin{table*}[htbp]
\centering
\renewcommand{\arraystretch}{1.6}
\caption{The summarized expressions and numerical values of the coupling constants in Eqs.~\eqref{eq:5}–\eqref{eq:8}. It is important to note that when calculating the values of these coupling constants, the binding momentum $\kappa$, the mass of the molecular pentaquark $m_\mathbb{P}$, the reduced mass $\mu$, and the mass of the constituent baryon $m_B$ must all be replaced with the corresponding values for each specific channel. For example, in the case of $g_\psi$, we have $\mu=m_{\Sigma_c}m_{\bar{D}}/(m_{\Sigma_c}+m_{\bar{D}})$, $m_B=m_{\Sigma_c}$, $m_\mathbb{P}=4311.9^{+6.8}_{-0.9}$ MeV~\cite{LHCb:2019kea}, and $\kappa=[2\mu(m_{\Sigma_c}+m_{\bar{D}}-m_\mathbb{P})]^{1/2}$. In the $\Sigma_c\bar{D}^\ast$ and $\Xi_c\bar{D}^\ast$ channels, the values marked with superscripts $\dagger$ and $\sharp$ correspond to the coupling constants for Case 1 [in Eqs.~\eqref{eq:PcCase1},~\eqref{eq:PcsCase1}] and Case 2 [in Eqs.~\eqref{eq:PcCase2},~\eqref{eq:PcsCase2}], respectively.\label{tab:gvalues}}
\setlength{\tabcolsep}{4.35mm}
{
\begin{tabular}{cccccc}
\hline 
\hline
$\Sigma_{c}\bar{D}$ & \multicolumn{2}{c}{$\Sigma_{c}\bar{D}^{\ast}$} & $\Xi_{c}\bar{D}$ & \multicolumn{2}{c}{$\Xi_{c}\bar{D}^{\ast}$}\tabularnewline
\hline 
$J=\frac{1}{2}$ & $J=\frac{1}{2}$ & $J=\frac{3}{2}$ & $J=\frac{1}{2}$ & $J=\frac{1}{2}$ & $J=\frac{3}{2}$\tabularnewline
\hline 
$g_{\psi}=2\sqrt{\frac{\pi\kappa m_{\mathbb{P}}}{\mu m_{B}}}$ & $\tilde{g}_{\psi}=2\sqrt{\frac{\pi\kappa m_{\mathbb{P}}}{3\mu m_{B}}}$ & $\tilde{g}_{\psi}^{\prime}=2\sqrt{\frac{\pi\kappa m_{\mathbb{P}}}{\mu m_{B}}}$ & $g_{\psi s}=2\sqrt{\frac{\pi\kappa m_{\mathbb{P}}}{\mu m_{B}}}$ & $\tilde{g}_{\psi s}=2\sqrt{\frac{\pi\kappa m_{\mathbb{P}}}{3\mu m_{B}}}$ & $\tilde{g}_{\psi s}^{\prime}=2\sqrt{\frac{\pi\kappa m_{\mathbb{P}}}{\mu m_{B}}}$\tabularnewline
\hline 
\multirow{2}{*}{$1.69_{-0.50}^{+0.04}$} & $\left(1.23_{-0.07}^{+0.06}\right)^{\dagger}$ & $\left(1.45_{-0.60}^{+0.12}\right)^{\dagger}$ & \multirow{2}{*}{$1.09_{-0.18}^{+0.16}$} & $\left(0.79_{-0.28}^{+0.17}\right)^{\dagger}$ & $\left(1.06_{-0.60}^{+0.30}\right)^{\dagger}$\tabularnewline
\cline{2-3}\cline{5-6}
 & $\left(0.84_{-0.33}^{+0.07}\right)^{\sharp}$ & $\left(2.12_{-0.11}^{+0.11}\right)^{\sharp}$ &  & $\left(0.61_{-0.40}^{+0.17}\right)^{\sharp}$ & $\left(1.37_{-0.50}^{+0.30}\right)^{\sharp}$\tabularnewline
\hline 
\hline 
\end{tabular}
}
\end{table*}

\subsection{Decay amplitudes and form factors}

\begin{figure*}[htbp]
\begin{centering}
    \scalebox{1.0}{\includegraphics[width=0.8\linewidth]{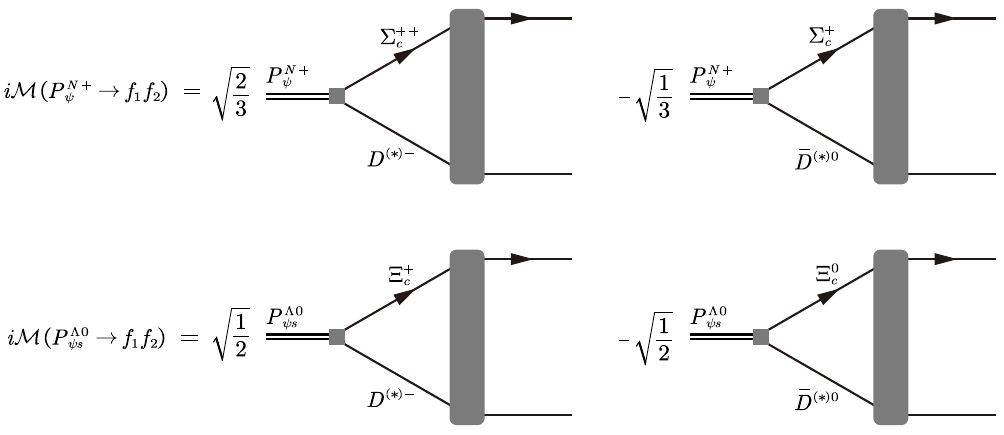}}
    \caption{An illustration of the Eqs.~\eqref{eq:iso12} and~\eqref{eq:iso0} in calculating the decay amplitudes, respectively.\label{fig:decayiso}}
\end{centering}
\end{figure*}

To obtain the decay amplitudes of \(P_\psi^N\) and \(P_{\psi s}^\Lambda\), we must account for their isospin wave functions. \(P_\psi^N\) and \(P_{\psi s}^\Lambda\) are states with isospins \(I = 1/2\) and \(I = 0\), respectively. Their forms in the two-particle basis are:
\begin{eqnarray}
|P_\psi^{N+}\rangle&=&\sqrt{\frac{2}{3}}|\Sigma_c^{++}D^{(\ast)-}\rangle-\sqrt{\frac{1}{3}}|\Sigma_c^{+}\bar{D}^{(\ast)0}\rangle,\label{eq:iso12}\\
|P_{\psi s}^{\Lambda0}\rangle&=&\sqrt{\frac{1}{2}}|\Xi_c^+ D^{(\ast)-}\rangle-\sqrt{\frac{1}{2}}|\Xi_c^0 \bar{D}^{(\ast)0}\rangle,\label{eq:iso0}
\end{eqnarray}
In Fig.~\ref{fig:decayiso}, we illustrate how the Eqs.~\eqref{eq:iso12} and~\eqref{eq:iso0} are reflected in the Feynman diagrams when calculating the decay amplitudes.

Consider the decays of \(P_\psi^N\) and \(P_{\psi s}^\Lambda\) into final states \(f_1\) and \(f_2\) via processes such as
\begin{eqnarray}
P_{\psi}^N/P_{\psi s}^\Lambda&\to& \Sigma_c/\Xi_c(p_1+q)\bar{D}^{(\ast)}(p_2-q)[M(q)]\nonumber \\ &\to&f_1(p_1)f_2(p_2),
\end{eqnarray}
where the quantities in parentheses denote the momentum of each particle, and \([M(q)]\) represents the transition of \(\Sigma_c \bar{D}^{(*)}/\Xi_c \bar{D}^{(*)}\) to \(f_1\) and \(f_2\) through the exchange of a meson \(M\) with momentum $q$. The vertices in Fig.~\ref{fig:decays} can be derived by expanding the Lagrangians provided in Sec.~\ref{sec:efflags}. 

The decay amplitudes corresponding to each Feynman diagram in Fig.~\ref{fig:decays} are given in detail in Appendix~\ref{sec:ap}.

The loop integrals presented in Appendix ~\ref{sec:ap} are generally divergent. To ensure convergence of the integrals, we introduce the following three commonly used types of form factors to regularize the loop integrals, namely by multiplying these form factors into the integrands. Detailed calculations can be found in Ref.~\cite{Deng:2024pep}.
\begin{eqnarray}
\text{Heaviside form factor}:\mathfrak{F}_1(|\bm{p}|)&=&\mathrm{\Theta}(\Lambda_1-|\bm{p}|),\\
\text{Gaussian form factor}:\mathfrak{F}_2(\bm{p}^2)&=&\mathrm{exp}\left(-\frac{\bm{p}^2}{\Lambda_2^2}\right),\\ 
\text{Multipole form factor}:\mathfrak{F}_3(p^2)&=&\left(\frac{m_M-\Lambda_3^2}{p^2-\Lambda_3^2}\right)^n,\label{eq:73}
\end{eqnarray}
where $\Lambda_i (i = 1, 2, 3)$ are cutoffs, and $m_M$ is the mass of the exchanged particle. Here, $\Lambda_3 = m_M + \alpha_\Lambda \Lambda_{\mathrm{QCD}}$, where $\alpha_\Lambda$ is a dimensionless parameter typically around $1$~\cite{Wang:2016qmz}.

The expression for calculating the decay width of \(P_\psi^N\) and \(P_{\psi s}^\Lambda\) into their corresponding final states \(f_1 f_2\) is given by:
\begin{eqnarray}
\Gamma_{f_1f_2} &=& \frac{1}{2J+1}\frac{|\bm{p}_1|}{8\pi m_{\mathbb{P}}^2}\overline{|\mathcal{M}_{f_1f_2}|^2},
\end{eqnarray}
where $J$ and $m_\mathbb{P}$ denote the spin and mass of the mother particle $P_\psi^N/P_{\psi s}^\Lambda$, respectively, and
\begin{eqnarray}
|\bm{p}_1|&=&\frac{\sqrt{\mathcal{K}(m_{\mathbb{P}}^2,m_{f_1}^2,m_{f_2}^2)}}{2m_{\mathbb{P}}},
\end{eqnarray}
with the triangle function
\begin{eqnarray}
\mathcal{K}(\alpha,\beta,\gamma)&=&\alpha ^2 + \beta ^2 + \gamma ^2-2 \alpha  \beta -2 \alpha  \gamma -2 \beta  \gamma.
\end{eqnarray}

\section{Numerical results and discussions}\label{sec:nums}

\subsection{$\Sigma_c\bar{D}$ system: $P_\psi^N(4312)$}

\begin{figure*}[htbp]
\begin{centering}
    \scalebox{1.0}{\includegraphics[width=\linewidth]{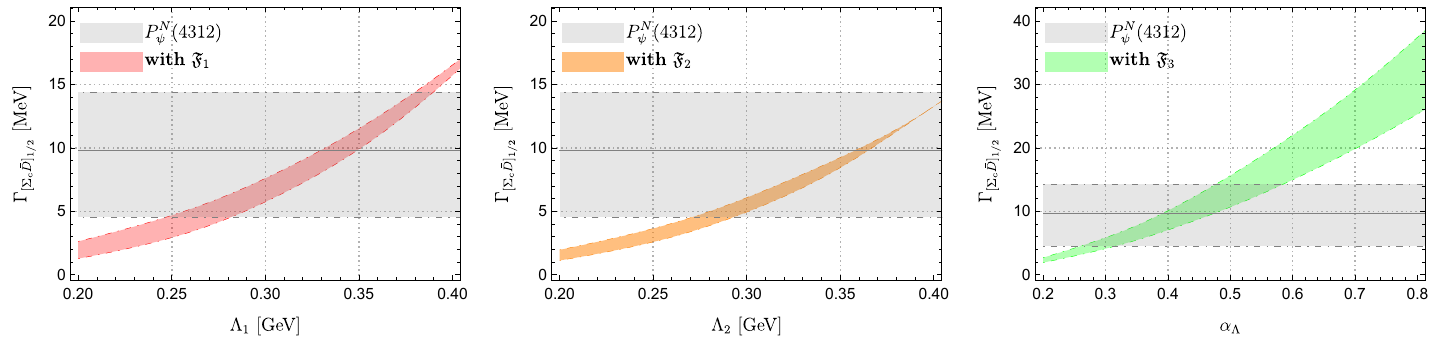}}
    \caption{The dependence of the total decay width of \(P_\psi^N(4312)\) on the cutoff parameters, where the horizontal shaded band represents the width measured by the LHCb Collaboration~\cite{LHCb:2019kea}, and the band corresponding to our results originates from the range of coupling constants given in Table~\ref{tab:gvalues}.\label{fig:tot_Pc4312}}
\end{centering}
\end{figure*}

In Fig.~\ref{fig:tot_Pc4312}, we present the dependence of the total decay width of \(P_\psi^N(4312)\) on the cutoff parameters under three different form factors. It can be seen that the total width increases with the cutoff parameter. The results obtained using form factors $\mathfrak{F}_1$ and $\mathfrak{F}_2$ are quite similar, with the calculated widths matching the experimental range when the cutoffs are taken as $0.3–0.4$ GeV. When the multipole form factor $\mathfrak{F}_3$ is employed, the calculated results are consistent with experimental measurements for \(\alpha_\Lambda\) in the range of $0.3–0.5$. The small value of the cutoff is a natural consequence in the molecular picture, since the typical interaction radius \( R \) of a molecular state satisfies \( R \sim 1/\Lambda \). That is, the fact that the experimental total width can be reproduced with a small cutoff value indicates that \(P_\psi^N(4312)\) corresponds to a loosely bound molecular configuration, rather than a compact multiquark state.

In Table~\ref{tab:parPc4312}, we present the ranges of partial decay widths of \(P_\psi^N(4312)\) into different channels. It can be seen that its dominant decay channel is \(\Lambda_c\bar{D}^*\), and the results from Refs.~\cite{Lin:2019qiv,Li:2023zag,Dong:2020nwk} are consistent with our calculations. Furthermore, the contribution of the \(\Lambda_c\bar{D}\) channel vanishes under heavy quark spin symmetry. The reason is that the two-particle bases \(\Sigma_c\bar{D}\) and \(\Lambda_c\bar{D}\) can be expanded in terms of the heavy quark spin symmetry (HQSS) basis as follows, 
\begin{align}
|[\Sigma_c\bar{D}]_{1/2}\rangle &= \frac{1}{2}|0_h\otimes\frac{1}{2}_\ell\rangle-\frac{1}{2\sqrt{3}}|1_h\otimes\frac{1}{2}_\ell\rangle\nonumber\\
&\quad+\sqrt{\frac{2}{3}}|1_h\otimes\frac{3}{2}_\ell\rangle, \\
|[\Lambda_c\bar{D}]_{1/2}\rangle &= \frac{1}{2}|0_h\otimes\frac{1}{2}_\ell\rangle+\frac{\sqrt{3}}{2}|1_h\otimes\frac{1}{2}_\ell\rangle,
\end{align}
where \(j_h\) and \(j_\ell\) denote the heavy spin and light spin, respectively. Both are conserved quantities under heavy quark spin symmetry, thus the transition amplitude satisfies the following relation, 
\begin{eqnarray}\label{eq:hqss1}
\langle [\Lambda_c\bar{D}]_{1/2}|\mathcal{L}_{\mathrm{HQSS}}|[\Sigma_c\bar{D}]_{1/2}\rangle=0,
\end{eqnarray}
with \(\mathcal{L}_{\mathrm{HQSS}}\) representing the interaction Lagrangian consistent with heavy quark spin symmetry. Since the interaction dynamics are dominated by the light spin degrees of freedom, it follows that $\langle\mathcal{L}_{\mathrm{HQSS}}\rangle_{0_h\otimes\frac{1}{2}_\ell} = \langle\mathcal{L}_{\mathrm{HQSS}}\rangle_{1_h\otimes\frac{1}{2}_\ell}$.

\begin{table*}[htbp]
\centering
\renewcommand{\arraystretch}{1.5}
 \caption{{\raggedright}The ranges of the partial decay widths of \(P_\psi^N(4312)\) obtained using different form factors (in units of MeV), along with comparisons to results from other studies in the literature. A dash indicates that the contribution of the corresponding channel is negligible.	\label{tab:parPc4312}}
\setlength{\tabcolsep}{2.75mm}
{
	\begin{tabular}{cccccccc}
		\hline	\hline
\multicolumn{1}{c}{Decay channels}	& With $\mathfrak{F}_1$ &   With $\mathfrak{F}_2$ &   With $\mathfrak{F}_3$ &Ref.~\cite{Lin:2019qiv} &Ref.~\cite{Li:2023zag}&Ref.~\cite{Dong:2020nwk} & LHCb~\cite{LHCb:2019kea}\\
		\hline		
$P_{\psi}^N(4312)\to\Lambda_c\bar{D}$  &  $-$ & $-$ & $-$& $0.3$& $0.026^{-0.05}_{+0.06}$& $-$& \\ 
$P_{\psi}^N(4312)\to\Lambda_c\bar{D}^\ast$  &  $4.6-14.3$ &$4.5-14.4$ & $4.5-14.2$& $10.7$& $8.8^{-1.6}_{+1.9}$& $8.36^{+3.68}_{-3.01}$&\\
$P_{\psi}^N(4312)\to pJ/\psi$  &  $0.03-0.09$ & $0.03-0.09$ & $0.04-0.2$& $10^{-3}- 0.1$& $0.17^{-0.04}_{+0.04}$&$0.0448^{+0.0197}_{-0.0161}$& \\
$P_{\psi}^N(4312)\to p\eta_c$  &  $-$ & $-$ & $0.002-0.01$&$10^{-2}- 0.4$& $0.085^{-0.016}_{+0.018}$&$0.0892^{+0.0392}_{-0.0321}$& \\ \hline  
\multicolumn{1}{c}{Total width}  &$4.6-14.4$  &$4.5-14.5$ &$4.5-14.2$ & $13.2$&$9.08^{-1.71}_{+2.02}$ &$8.49^{+3.74}_{-3.06}$ &$9.8\pm2.7^{+3.7}_{-4.5}$ \\		
\hline	\hline
	\end{tabular}
}
\end{table*}

In Table~\ref{tab:fraPc4312}, we also list the branching fractions of \(P_\psi^N(4312)\) in different decay channels. It can be seen that although the partial and total decay widths are rather sensitive to the cutoff, the branching fractions are rather insensitive to it. Therefore, branching fractions can, to some extent, serve as a quantity with weak model dependence for probing the internal structure of \(P_\psi^N(4312)\).

\begin{table}[htbp]
\renewcommand{\arraystretch}{1.5}
	\centering
	\caption{The branching fractions of \({P_{\psi}^N}(4312)\) within different form factors.	\label{tab:fraPc4312}}
\setlength{\tabcolsep}{3.8mm}
{
	\begin{tabular}{cccc}
		\hline	\hline
		Form factors&  With $\mathfrak{F}_1$ & With $\mathfrak{F}_2$ & With $\mathfrak{F}_3$ \\
		\hline
		$\Gamma_{\Lambda_c\bar{D}}/\Gamma_{{P_{\psi}^N}(4312)}$  &  $-$ & $-$ & $-$ \\ 
        $\Gamma_{\Lambda_c\bar{D}^\ast}/\Gamma_{{P_{\psi}^N}(4312)}$  &  $0.99$ & $0.99$ & $0.99$ \\
        $\Gamma_{pJ/\psi}/\Gamma_{{P_{\psi}^N}(4312)}$  & $0.01$ &$0.01$ & $0.01$ \\
        $\Gamma_{p\eta_c}/\Gamma_{{P_{\psi}^N}(4312)}$  &  $-$ & $-$ & $0.001$ \\
		\hline	\hline
	\end{tabular}
}
\end{table}

Since \(P_\psi^N(4312)\) has relatively few decay channels, with only \(\Lambda_c\bar{D}^\ast\) making a dominant contribution, its total decay width can serve as a clean window to constrain the range of cutoff parameters. Based on the overlap between the calculated results and the experimental range in Fig.~\ref{fig:tot_Pc4312}, the ranges for \(\Lambda_1\), \(\Lambda_2\), and \(\alpha_\Lambda\) are determined to be:  
\(\Lambda_1 \in [0.28,0.38]\) GeV,  
\(\Lambda_2 \in [0.29,0.41]\) GeV,  
and \(\alpha_\Lambda \in [0.32,0.48]\).  
For other systems, we will adopt these cutoff values to predict their partial decay widths and corresponding branching fractions.

\subsection{$\Sigma_c\bar{D}^\ast$ system: $P_\psi^N(4440)$ and $P_\psi^N(4457)$}

In the molecular framework, the total angular momentum of the \(\Sigma_c \bar{D}^*\) system can take values of \(J = 1/2\) and \(J = 3/2\). Since the spin quantum numbers of \(P_\psi^N(4440)\) and \(P_\psi^N(4457)\) have not yet been experimentally determined, their spins are subject to the following two possibilities:

\begin{eqnarray}
\textbf{Case 1}&&J=\frac{1}{2}: P_\psi^N(4440),~~ J=\frac{3}{2}: P_\psi^N(4457),\label{eq:PcCase1}\\
\textbf{Case 2}&&J=\frac{1}{2}: P_\psi^N(4457),~~ J=\frac{3}{2}: P_\psi^N(4440).\label{eq:PcCase2}
\end{eqnarray}

\begin{figure*}[htbp]
\begin{centering}
    \scalebox{1.0}{\includegraphics[width=\linewidth]{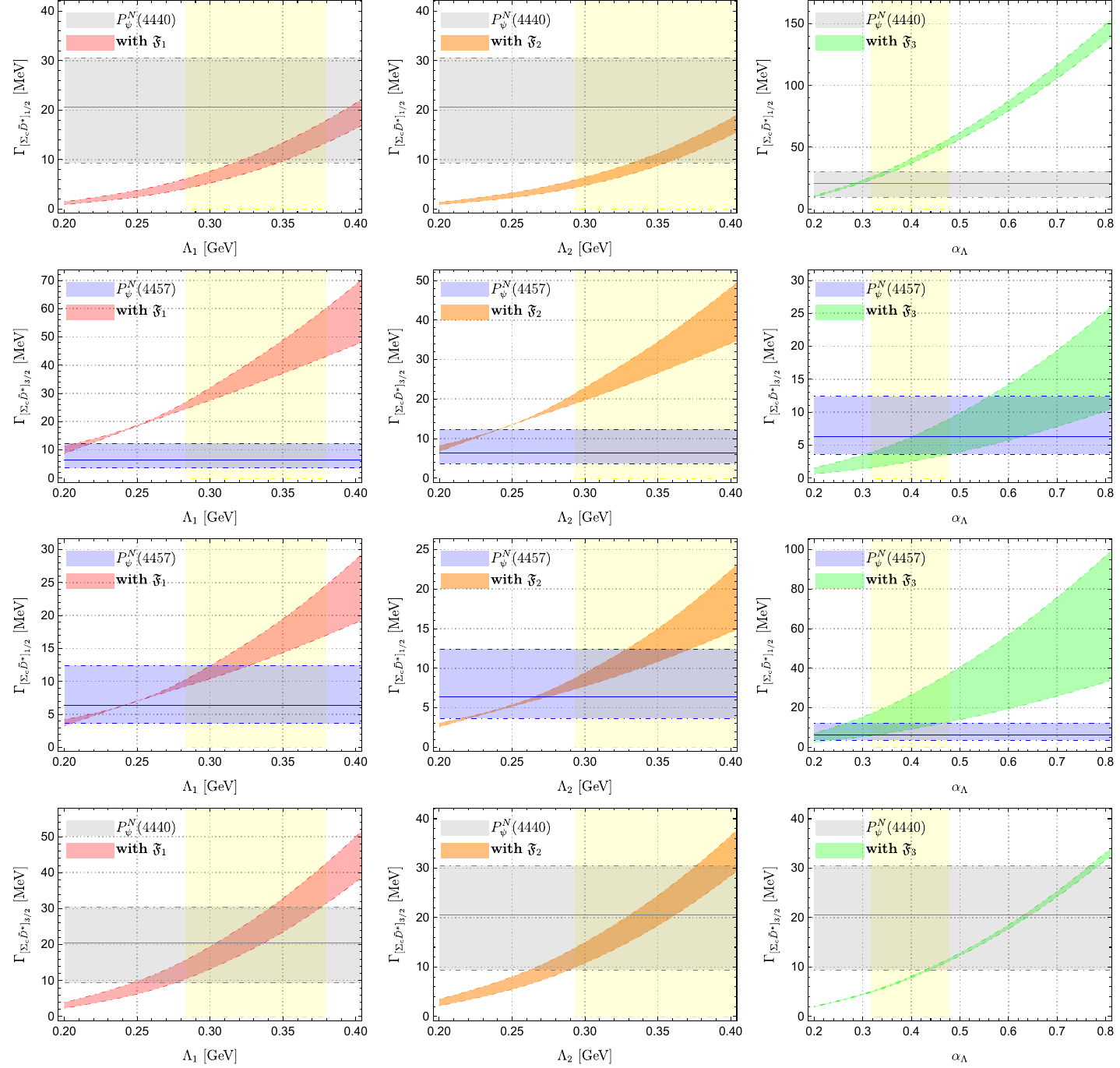}}
    \caption{The dependence of the total decay widths of \(P_\psi^N(4440)\) and \(P_\psi^N(4457)\) on the cutoff parameters, where the first two rows correspond to the results for Case 1, and the latter two rows correspond to those for Case 2. The vertical yellow band indicates the constrained range of the corresponding cutoff parameters, with values \(\Lambda_1 \in [0.28, 0.38]\) GeV, \(\Lambda_2 \in [0.29, 0.41]\) GeV, and \(\alpha_\Lambda \in [0.32, 0.48]\). The other notations used in the figure are the same as those in Fig.~\ref{fig:tot_Pc4312}.\label{fig:tot_Pc4440_Pc4457}}
\end{centering}
\end{figure*}

In Fig.~\ref{fig:tot_Pc4440_Pc4457}, we illustrate the dependence of the total decay widths of \(P_\psi^N(4440)\) and \(P_\psi^N(4457)\) on the cutoff parameters for both Case 1 and Case 2. It can be seen that the overall behavior is similar to that in Fig.~\ref{fig:tot_Pc4312}. Within the molecular picture, the interactions in the \(\Sigma_c\bar{D}\) and \(\Sigma_c\bar{D}^*\) systems are closely related~\cite{Meng:2019ilv,Wang:2019ato}. Therefore, we use the experimental width of \(P_\psi^N(4312)\) to constrain the range of cutoff parameters (indicated by the vertical yellow band in Fig.~\ref{fig:tot_Pc4440_Pc4457}), and employ this constraint to study \(P_\psi^N(4440)\) and \(P_\psi^N(4457)\). From Fig.~\ref{fig:tot_Pc4440_Pc4457}, it can be observed that in Case 1, the width of \(P_\psi^N(4440)\) overlaps with the experimental value within the constrained cutoff range. However, for \(P_\psi^N(4457)\), when using form factors $\mathfrak{F}_1$ and $\mathfrak{F}_2$, its width significantly exceeds the experimental range. In Case 2, regardless of which form factor is used, the widths of both \(P_\psi^N(4440)\) and \(P_\psi^N(4457)\) overlap with the experimental values within the constrained cutoff range. From this perspective, it is more likely that the spin quantum numbers of \(P_\psi^N(4440)\) and \(P_\psi^N(4457)\) correspond to those in Case 2. The conclusions in Refs.~\cite{Du:2019pij,Liu:2019zvb} are consistent with ours.

\begin{table*}
\centering
\renewcommand{\arraystretch}{1.5}
\caption{The ranges of partial decay widths for \(P_\psi^N(4440)\) and \(P_\psi^N(4457)\) in Case 1 [in Eq.~\eqref{eq:PcCase1}] and Case 2 [in Eq.~\eqref{eq:PcCase2}] using different form factors (in units of MeV), where we adopt \(\Lambda_1 \in [0.28, 0.38]\) GeV, \(\Lambda_2 \in [0.29, 0.41]\) GeV, and \(\alpha_\Lambda \in [0.32, 0.48]\) to present the ranges of each partial width.	\label{tab:parPc4440_Pc4457}}
\setlength{\tabcolsep}{1.75mm}
{
	\begin{tabular}{ccccccccc}
		\hline
        \hline
		\multicolumn{1}{c}{} & \multicolumn{1}{c}{Form factors} 	&   With $\mathfrak{F}_1$ &   With $\mathfrak{F}_2$ &   With $\mathfrak{F}_3$ &Ref.~\cite{Li:2025ejt}&Ref.~\cite{Lin:2019qiv} & Ref.~\cite{Yang:2024nss} &LHCb\cite{LHCb:2019kea}\\
		\hline
		
		\multirow{14}{*}{\textbf{Case 1}} 
		& $P_{\psi}^N(4440)\to\Lambda_c\bar{D}$  & $0.4-2.3$ & $0.5-2.8$ & $10.9-26.8$& $4.6$& $5.6$&$5.98$& \\ 
		& $P_{\psi}^N(4440)\to\Lambda_c\bar{D}^\ast$  &  $1.8-7.9$ & $1.8-9.2$ &$7.7-18.9$& $7.6$&$13.9$& $-$&  \\  
		& $P_{\psi}^N(4440)\to\Sigma_c\bar{D}$ &  $0.9-3.5$ & $0.9-3.8$ & $2.6-6.2$& $5.2\times10^{-2}$& $3.4$& $10.43$&\\ 
		& $P_{\psi}^N(4440)\to\Sigma^\ast_c\bar{D}$ &  $1.0-3.9$ & $1.0-3.9$ & $2.0-4.7$& $3.6\times10^{-1}$& $0.8$& $-$&\\
		& $P_{\psi}^N(4440)\to pJ/\psi$ &  $0.04-0.2$ & $0.04-0.2$ & $0.05-0.1$& $1.7\times10^{-2}$& $0.03$& $4.10$&  \\
		& $P_{\psi}^N(4440)\to p\eta_c$  &  $0.04-0.2$ & $0.04-0.2$ & $0.2-0.4$& $3.5\times10^{-1}$& $3\times10^{-4}$& -- \\ 
		
		\cmidrule{2-9} 	&\multicolumn{1}{c}{Total width}  & $4.1-17.9$ & $4.3-20.02$ & $23.5-51.1$& $13.0$& $23.7$&$20.52$ & \scalebox{0.95}{$20.6\pm4.9^{+8.7}_{-10.1}$} \\
		\cmidrule{2-9}
		
		& $P_{\psi}^N(4457)\to\Lambda_c\bar{D}$  &  $15.1-38.2$ &$11.5-30.9$ &$0.3-1.4$& $16.7$& $1.5$&$2.15$ \\
		& $P_{\psi}^N(4457)\to\Lambda_c\bar{D}^\ast$  & $3.4-11.0$ & $2.9-11.7$ & $1.07-6.8$& $8.0\times10^{-1}$& $6.1$&$-$   \\  
		& $P_{\psi}^N(4457)\to\Sigma_c\bar{D}$  & $3.3-6.8$ & $2.5-5.2$ & $0.09-0.3$&$8.1\times10^{-2}$&$1.0$&$4.11$   \\ 
		& $P_{\psi}^N(4457)\to\Sigma^\ast_c\bar{D}$  & $2.2-4.2$ &$1.6-3.2$ & $0.1-0.4$&$7.4\times10^{-2}$& $6.2$& $-$ \\
		& $P_{\psi}^N(4457)\to pJ/\psi$   &  $0.03-0.09$ & $0.03-0.08$ & $0.007-0.04$&$5.8\times10^{-3}$& $0.01$& $1.52$&  \\
		& $P_{\psi}^N(4457)\to p\eta_c$  &  $0.02-0.04$ &$0.01-0.04$ & $0.002-0.005$& $1.5\times10^{-1}$&$-$&$-$&  \\ 
		
		\cmidrule{2-9}
		&\multicolumn{1}{c}{Total width}  & $24.4-60.3$ & $18.8-51.2$ & $1.6-9.04$& $18.5$& $14.7$&$7.78$ &$6.4\pm2.0^{+5.7}_{-1.9}$ \\
		\hline
		\multirow{15}{*}{\textbf{Case 2}}
		& $P_{\psi}^N(4440)\to\Lambda_c\bar{D}$&  $7.1-27.8$ & $6.5-23.7$ & $0.6-1.5$& $20.0$& $1.7$&$4.53$ \\
		& $P_{\psi}^N(4440)\to\Lambda_c\bar{D}^\ast$  &   $1.8-8.6$ & $2.2-10.5$ & $3.9-9.8$&  $8.8\times10^{-1}$& $6.2$&$-$\\
		& $P_{\psi}^N(4440)\to\Sigma_c\bar{D}$  & $1.05-4.2$ & $0.9-3.3$ & $0.09-0.3$& $7.5\times10^{-2}$& $0.5$&$5.35$  \\
		& $P_{\psi}^N(4440)\to\Sigma^\ast_c\bar{D}$ & $0.4-2.1$ & $0.4-1.6$ & $0.1-0.3$&  $6.4\times10^{-1}$& $5.4$&$-$\\
		& $P_{\psi}^N(4440)\to pJ/\psi$  &  $0.01-0.06$ & $0.01-0.07$ & $0.03-0.06$& $6.8\times10^{-3}$& $0.02$& $4.10$\\
		& $P_{\psi}^N(4440)\to p\eta_c$  &  $0.008-0.03$ & $0.008-0.03$ & $0.003-0.007$& $1.9\times10^{-1}$& $4\times10^{-5}$&$-$   \\
		
		\cmidrule{2-9}
		& \multicolumn{1}{c}{Total width}&$10.5-42.7$ & $9.9-39.3$ & $4.8-11.8$& $21.8$& $13.9$&$13.98$ &\scalebox{0.95}{$20.6\pm4.9^{+8.7}_{-10.1}$}\\
		\cmidrule{2-9}
		
		& $P_{\psi}^N(4457)\to\Lambda_c\bar{D}$  & $ 1.3-3.5$ & $1.08-3.7$ & $2.7-17.2$& $4.3$& $3.8$&$2.47$\\
		& $P_{\psi}^N(4457)\to\Lambda_c\bar{D}^\ast$    &  $4.9-11.8$ & $3.8-11.5$ & $2.0-12.6$&$7.8$& $12.5$& $-$  \\
		& $P_{\psi}^N(4457)\to\Sigma_c\bar{D}$          & $ 1.2-4.1$ & $1.0-4.0$ & $0.6-4.0$& $5.5\times10^{-2}$& $2.6$& $5.60$ \\
		& $P_{\psi}^N(4457)\to\Sigma^\ast_c\bar{D}$      & $1.7-4.9$ & $1.3-4.5$ & $0.5-3.3$&$4.7\times10^{-1}$& $1.9$& $-$ \\
		& $P_{\psi}^N(4457)\to pJ/\psi$                 &  $0.05-0.2$ & $0.04-0.2$ & $0.01-0.03$& $1.6\times10^{-2}$& $0.02$& $1.52$  \\
		& $P_{\psi}^N(4457)\to p\eta_c$                 &  $0.06-0.2$ & $0.05-0.2$ & $0.03-0.2$& $3.2\times10^{-1}$&$-$& $-$ \\
		
		\cmidrule{2-9}
		& \multicolumn{1}{c}{Total width}  & $9.2-24.6$ & $7.3-24.1$ & $5.9-37.4$& $13.0$& $20.7$&$9.59$& $6.4\pm2.0^{+5.7}_{-1.9}$\\
\hline \hline
	\end{tabular}
}
\end{table*}

\begin{table*}[!htbp]
	\centering
\renewcommand{\arraystretch}{1.5}
	\caption{The ranges of branching fractions for \(P_\psi^N(4440)\) and \(P_\psi^N(4457)\) in Case 1 [in Eq.~\eqref{eq:PcCase1}] and Case 2 [in Eq.~\eqref{eq:PcCase2}] using different form factors.	\label{tab:fraPc4440_Pc4457}}
\setlength{\tabcolsep}{3.25mm}
{
\begin{tabular}{ccccccc}
	\hline
    \hline
	\multicolumn{1}{c}{} & \multicolumn{3}{c}{\textbf{Case 1}} & \multicolumn{3}{c}{\textbf{Case 2}} \\
	\cmidrule(lr){2-4} \cmidrule(lr){5-7}
	\multicolumn{1}{c}{Form factor} & With $\mathfrak{F}_1$ & With $\mathfrak{F}_2$ & With $\mathfrak{F}_3$ 
	& With $\mathfrak{F}_1$ & With $\mathfrak{F}_2$ & With $\mathfrak{F}_3$ \\
	\hline
		$\Gamma_{\Lambda_c\bar{D}}/\Gamma_{{P_{\psi}^N}(4440)}$  & $0.09-0.12$ &$0.1-0.13$ & $0.46-0.47$ &$0.65-0.68$&$0.61-0.68$&$0.12-0.13$\\ $\Gamma_{\Lambda_c\bar{D}^\ast}/\Gamma_{{P_{\psi}^N}(4440)}$  &  $0.43-0.48$ & $0.44-0.46$ & $0.33-0.34$&$0.15-0.22$&$0.17-0.28$&$0.82-0.83$ \\$\Gamma_{\Sigma_c\bar{D}}/\Gamma_{{P_{\psi}^N}(4440)}$  & $0.19-0.21$ & $0.20$ & $0.11$&$0.09-0.12$&$0.08-0.11$&$0.02$ \\$\Gamma_{\Sigma^\ast_c\bar{D}}/\Gamma_{{P_{\psi}^N}(4440)}$  & $0.22-0.23$ & $0.2-0.23$ &$0.08-0.09$&$0.04-0.06$&$0.03-0.05$&$0.02-0.03$ \\	$\Gamma_{pJ/\psi}/\Gamma_{{P_{\psi}^N}(4440)}$  &  $0.01$ &$0.01$ &$0.002$&$0.001-0.002$&$0.001-0.002$&$0.01$ \\ $\Gamma_{p\eta_c}/\Gamma_{{P_{\psi}^N}(4440)}$  &$0.01$ & $0.01$ & $0.01$&$0.001$&$0.001$&$0.001$ \\	
\hline
$\Gamma_{\Lambda_c\bar{D}}/\Gamma_{{P_{\psi}^N}(4457)}$  &  $0.62-0.63$ & $0.61-0.63$&$0.17-0.18$&$0.12-0.15$&$0.13-0.16$&$0.46$ \\$\Gamma_{\Lambda_c\bar{D}^\ast}/\Gamma_{{P_{\psi}^N}(4457)}$  &  $0.13-0.18$ & $0.14-0.21$ & $0.7-0.73$&$0.49-0.55$&$0.49-0.53$&$0.33-0.35$ \\	$\Gamma_{\Sigma_c\bar{D}^\ast}/\Gamma_{{P_{\psi}^N}(4457)}$  &  $0.12-0.14$ & $0.11-0.14$ & $0.05$&$0.13-0.16$&$0.14-0.16$&$0.1-0.11$ \\ $\Gamma_{\Sigma^\ast_c\bar{D}}/\Gamma_{{P_{\psi}^N}(4457)}$  & $0.08-0.1$ & $0.08-0.1$ & $0.06$&$0.19$&$0.18-0.19$&$0.09$ \\$\Gamma_{pJ/\psi}/\Gamma_{{P_{\psi}^N}(4457)}$  & $ 0.001-0.002$ &$0.001-0.002$ & $0.004-0.005$&$0.01$&$0.01$&$0.002$ \\$\Gamma_{p\eta_c}/\Gamma_{{P_{\psi}^N}(4457)}$  &  $0.001$ & $0.001$ & $0.001$&$0.01$&$0.01$&$0.006$ \\
    \hline
    \hline
	\end{tabular}
}
\end{table*}

In Tables~\ref{tab:parPc4440_Pc4457} and \ref{tab:fraPc4440_Pc4457}, we present the partial decay widths and branching fractions of \(P_\psi^N(4440)\) and \(P_\psi^N(4457)\) for both Case 1 and Case 2, respectively. From Table~\ref{tab:fraPc4440_Pc4457}, it can be seen that the branching fractions are insensitive to the variations of the cutoff. When using form factors $\mathfrak{F}_1$ and $\mathfrak{F}_2$, the decay widths (and corresponding branching fractions) of \(P_\psi^N(4440)\) and \(P_\psi^N(4457)\) into the \(\Lambda_c\bar{D}\) and \(\Lambda_c\bar{D}^*\) channels differ significantly between Case 1 and Case 2. The relative magnitudes are summarized as follows:
\begin{eqnarray}\label{eq:widthrela}
\mathfrak{F}_{1}(\mathfrak{F}_{2})~\begin{cases}
\textrm{Case 1}\begin{cases}
P_{\psi}^{N}(4440): & \Gamma_{\Lambda_{c}\bar{D}}<\Gamma_{\Lambda_{c}\bar{D}^{\ast}}\\
P_{\psi}^{N}(4457): & \Gamma_{\Lambda_{c}\bar{D}}>\Gamma_{\Lambda_{c}\bar{D}^{\ast}}
\end{cases}\\
\textrm{Case 2}\begin{cases}
P_{\psi}^{N}(4440): & \Gamma_{\Lambda_{c}\bar{D}}>\Gamma_{\Lambda_{c}\bar{D}^{\ast}}\\
P_{\psi}^{N}(4457): & \Gamma_{\Lambda_{c}\bar{D}}<\Gamma_{\Lambda_{c}\bar{D}^{\ast}}
\end{cases}
\end{cases}.
\end{eqnarray}
However, when using the form factor $\mathfrak{F}_3$, the size relationship is completely opposite to that in Eq.~\eqref{eq:widthrela}. Although it is difficult to determine which form factor is more physically reasonable, it can be seen from Fig.~\ref{fig:tot_Pc4440_Pc4457} that, within the constrained range, the band representing the width of \(P_\psi^N(4440)\) obtained with $\mathfrak{F}_3$ has a tiny overlap with the experimental measurement, compared to those obtained with $\mathfrak{F}_1$ and $\mathfrak{F}_2$. This may suggest that the results calculated with form factors $\mathfrak{F}_1$ and $\mathfrak{F}_2$ are somewhat more stable. Therefore, future experiments may determine the spin configurations of \(P_\psi^N(4440)\) and \(P_\psi^N(4457)\) by measuring their partial decay widths into \(\Lambda_c\bar{D}\) and \(\Lambda_c\bar{D}^*\).

\subsection{$\Xi_c\bar{D}$ system: $P_{\psi s}^\Lambda(4338)$}

The experimentally measured mass of \(P_{\psi s}^\Lambda(4338)\) lies above the \(\Xi_c \bar{D}\) threshold~\cite{LHCb:2022ogu}. In our previous work~\cite{Meng:2022wgl}, we combined the coupled-channel approach with the Lippmann–Schwinger equation to study the lineshape of the \(J/\psi\Lambda\) invariant mass distribution, and pointed out that for such a near-threshold state, using a Breit–Wigner parametrization might lead to a misidentification of its pole position (i.e., its physical mass). In the present work, we treat \(P_{\psi s}^\Lambda(4338)\) as a bound state of \(\Xi_c \bar{D}\), and adopt its mass from Table IV of Ref.~\cite{Wang:2023eng}, namely 
\begin{eqnarray}
m_{[\Xi_c\bar{D}]_{1/2}} = 4334.8^{+0.8}_{-1.1} \textrm{ MeV}.
\end{eqnarray}

\begin{figure*}[htbp]
\begin{centering}
    \scalebox{1.0}{\includegraphics[width=\linewidth]{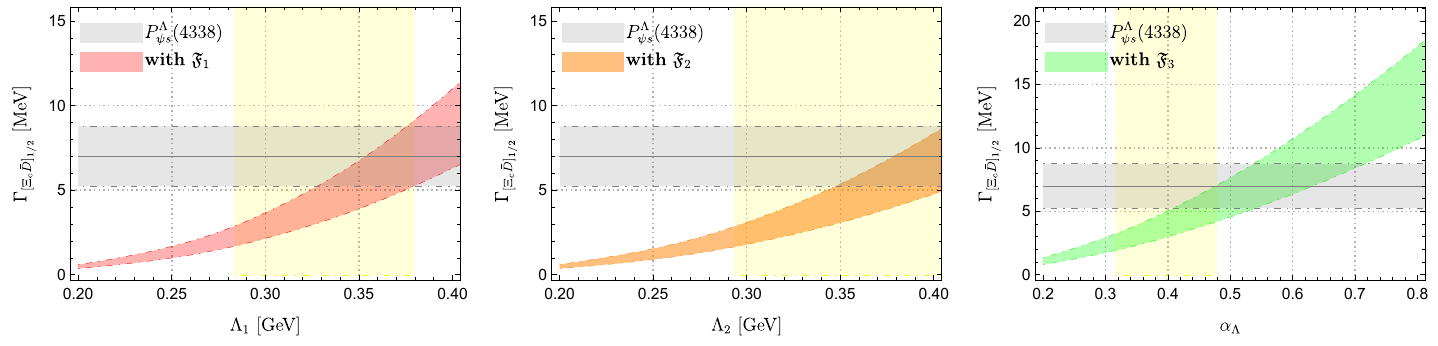}}
    \caption{The dependence of the total decay width of \(P_{\psi s}^\Lambda(4338)\) on the cutoff parameters, where the notations in the figure are the same as those in Fig.~\ref{fig:tot_Pc4440_Pc4457}.\label{fig:tot_Pcs4338}}
\end{centering}
\end{figure*}

In Fig.~\ref{fig:tot_Pcs4338}, we present the dependence of the total decay width of \(P_{\psi s}^\Lambda(4338)\) on the cutoff parameters. It can be seen that, within the constrained cutoff range, the theoretical results exhibit substantial overlap with the experimental measurements. From this perspective, our results support the molecular interpretation of \(P_{\psi s}^\Lambda(4338)\).

In Tables~\ref{tab:parPcs4338} and~\ref{tab:fraPcs4338}, we list the partial decay widths and corresponding branching fractions of \(P_{\psi s}^\Lambda(4338)\), respectively. It can be observed that its strong decay behavior is very similar to that of \(P_\psi^N(4312)\): the decay widths into hidden-charm channels (such as \(J/\psi\Lambda\) and \(\eta_c\Lambda\)) are significantly smaller than those into open-charm channels (e.g., \(\Lambda_c\bar{D}_s\)). Regardless of which form factor is used in the calculation, \(\Lambda_c\bar{D}_s\) remains the dominant decay channel.

\begin{table*}[htbp]
\renewcommand{\arraystretch}{1.5}
	\centering	\caption{The ranges of the partial decay widths of \(P_{\psi s}^\Lambda(4338)\) obtained using different form factors (in units of MeV).	\label{tab:parPcs4338}}
\setlength{\tabcolsep}{4.3mm}
{
	\begin{tabular}{ccccccc}
		\hline	\hline
		\multicolumn{1}{c}{Decay channels}	&  With $\mathfrak{F}_1$ &  With $\mathfrak{F}_2$ &  With $\mathfrak{F}_3$ & Ref.~\cite{Ortega:2022uyu}&Ref.~\cite{Shen:2019evi} & LHCb~\cite{LHCb:2022ogu}\\
		\hline
		$P_{\psi s}^\Lambda(4338)\to\Lambda_c\bar{D}_s$  &  $1.6-8.8$ & $1.6-8.7$ & $1.7-6.2$& $11.0$& $-$&  \\
        $P_{\psi s}^\Lambda(4338)\to \Lambda J/\psi$  &  $0.09-0.3$ & $0.08-0.3$ & $0.2-0.8$&$0.6$& $-$&  \\
        $P_{\psi s}^\Lambda(4338)\to\Lambda\eta_c$  &  $0.01-0.04$ & $0.001-0.05$ & $0.01-0.06$& $1.2$& $1.961$&  \\ 
        \hline \multicolumn{1}{c}{Total width}  & $1.7-9.2$ & $1.7-9.1$ & $1.9-7.1$& $12.8$ & $8.666$ & $7\pm1.2\pm1.3$ \\
		\hline	\hline
	\end{tabular}
}
\end{table*}

\begin{table}[htbp]
\renewcommand{\arraystretch}{1.5}
	\centering
	\caption{The branching fractions of \({P_{\psi s}^\Lambda}(4338)\) within different form factors.	\label{tab:fraPcs4338}}
	\begin{tabular}{cccc}
		\hline	\hline
		Form factor&  With $\mathfrak{F}_1$ & With $\mathfrak{F}_2$ & With $\mathfrak{F}_3$ \\
		\hline
		$\Gamma_{\Lambda_c\bar{D}_s}/\Gamma_{{P_{\psi s}^\Lambda}(4338)}$  &  $0.92-0.96$ & $0.93-0.96$ & $0.87-0.88$ \\ 
        $\Gamma_{\Lambda J/\psi}/\Gamma_{{P_{\psi s}^\Lambda}(4338)}$  & $0.03-0.08$ & $0.03-0.06$ & $0.11-0.12$ \\
        $\Gamma_{\Lambda \eta_c}/\Gamma_{{P_{\psi s}^\Lambda}(4338)}$  & $0.004-0.007$ & $0.005-0.007$ & $0.006-0.009$ \\
		\hline	\hline
	\end{tabular}
\end{table}

\subsection{$\Xi_c\bar{D}^\ast$ system: $P_{\psi s}^\Lambda(4459)$}

Just like the \(\Sigma_c\bar{D}^*\) system, the \(\Xi_c\bar{D}^*\) system can also have total angular momentum \(J = 1/2\) and \(J = 3/2\). The recently observed \(P_{\psi s}^\Lambda(4459)\) by the LHCb experiment~\cite{LHCb:2020jpq} is likely a candidate for a \(\Xi_c\bar{D}^*\) bound state. However, the current data are insufficient to determine whether its \(J^P\) quantum numbers are \(1/2^-\) or \(3/2^-\), or whether the experimental lineshape actually receives contributions from both \(1/2^-\) and \(3/2^-\) states. For the masses of the \(J = 1/2\) and \(J = 3/2\) states in the \(\Xi_c\bar{D}^*\) system, we also adopt the results from Ref.~\cite{Wang:2023eng}. It should be noted that the masses listed in Table IV of Ref.~\cite{Wang:2023eng} correspond to the scenario where the spins of \(P_c(4440)\) and \(P_c(4457)\) are assigned as in Case 1. Hence, we also need to compute the bound-state masses of the \(\Xi_c\bar{D}^*\) system for the spin assignments in Case 2. Following the approach of Ref.~\cite{Wang:2023eng}, we refitted the low-energy coupling constants \(c_s\) and \(c_a\) [see Eqs.~(1)–(4) in Ref.~\cite{Wang:2023eng}] for Case 2. We find that \(c_s\) remains almost unchanged, while \(c_a\) changes sign, which approximately leads to an interchange of the masses of the \(J = 1/2\) and \(J = 3/2\) states in the \(\Xi_c\bar{D}^*\) system. Consequently, for the \(\Xi_c\bar{D}^*\) system, we also consider the following two cases:
\begin{eqnarray}
\textbf{Case 1}&&\begin{cases}\label{eq:PcsCase1}
m_{[\Xi_{c}\bar{D}^{\ast}]_{1/2}}=4473.9_{-4.5}^{+3.1}\textrm{ MeV}\\
m_{[\Xi_{c}\bar{D}^{\ast}]_{3/2}}=4476.3_{-2.3}^{+1.3}\textrm{ MeV}
\end{cases},
\\
\textbf{Case 2}&&\begin{cases}\label{eq:PcsCase2}
m_{[\Xi_{c}\bar{D}^{\ast}]_{1/2}}=4476.3_{-2.3}^{+1.3}\textrm{ MeV}\\
m_{[\Xi_{c}\bar{D}^{\ast}]_{3/2}}=4473.9_{-4.5}^{+3.1}\textrm{ MeV}
\end{cases}.
\end{eqnarray}

\begin{figure*}[htbp]
\begin{centering}
    \scalebox{1.0}{\includegraphics[width=\linewidth]{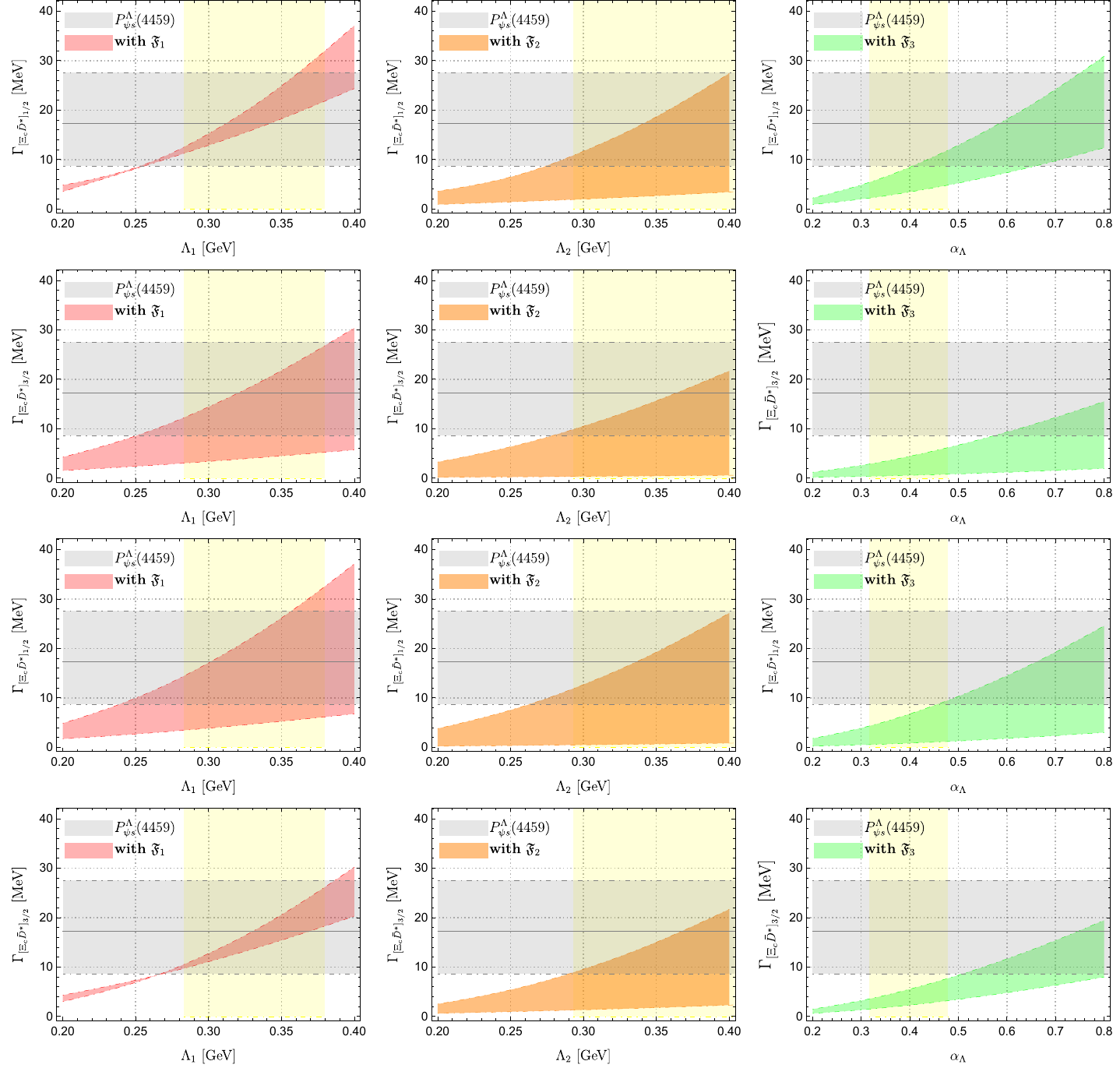}}
    \caption{The dependence of the decay widths of the two bound states in the \(\Xi_c\bar{D}^*\) system with \(J = 1/2\) and \(J = 3/2\) on the cutoff parameters. The figures in the first two rows correspond to the results of Case 1 in Eq.~\eqref{eq:PcsCase1}, while the last two rows correspond to the results of Case 2 in Eq.~\eqref{eq:PcsCase2}. The horizontal gray bands represent the width range of \(P_{\psi s}^\Lambda(4459)\) measured by LHCb~\cite{LHCb:2020jpq}. Other notations are the same as those in Fig.~\ref{fig:tot_Pc4440_Pc4457}.\label{fig:tot_Pcs4459}}
\end{centering}
\end{figure*}

In Fig.~\ref{fig:tot_Pcs4459}, we illustrate the dependence of the decay widths of the bound states in the \(\Xi_c\bar{D}^*\) system on the cutoff parameters for both Case 1 and Case 2. Taking the width of \(P_{\psi s}^\Lambda(4459)\) measured by LHCb as a reference~\cite{LHCb:2020jpq}, it can be read that, when using form factors $\mathfrak{F}_1$ and $\mathfrak{F}_2$, the calculated results overlap significantly with the experimental range. However, when form factor $\mathfrak{F}_3$ is used, the overlap is either very small or nonexistent. Due to the lack of experimental data, it is currently difficult to determine the \(J^P\) quantum numbers of \(P_{\psi s}^\Lambda(4459)\). Nevertheless, within the molecular picture, the \(\Xi_c\bar{D}^*\) system is very similar to the \(\Sigma_c\bar{D}^*\) system, both having two S-wave bound states with \(J = 1/2\) and \(J = 3/2\). Therefore, it is highly plausible that the experimentally observed lineshape of \(P_{\psi s}^\Lambda(4459)\)~\cite{LHCb:2020jpq} also contains two substructures, just as \(P_{\psi}^N(4450)\) was resolved into \(P_{\psi}^N(4440)\) and \(P_{\psi}^N(4457)\)~\cite{LHCb:2019kea}. From the experimental information on the \(P_{\psi}^N\) states, we find that \(\Gamma[P_{\psi}^N(4440)] > \Gamma[P_{\psi}^N(4457)]\), i.e., in the spin doublet of the \(\Sigma_c\bar{D}^*\) system, one state is broader and the other narrower. In the present calculation, our results support the Case 2 in Eq.~\eqref{eq:PcCase2}, where \(\Gamma_{[\Sigma_c\bar{D}^*]_{3/2}} > \Gamma_{[\Sigma_c\bar{D}^*]_{1/2}}\). Hence, it can be inferred that the decay widths of the molecular states in the \(\Xi_c\bar{D}^*\) system may also follow a similar pattern, namely \(\Gamma_{[\Xi_c\bar{D}^*]_{3/2}} > \Gamma_{[\Xi_c\bar{D}^*]_{1/2}}\).

In Tables~\ref{tab:parPcs4459} and~\ref{tab:fraPcs4459}, we present the partial decay widths and branching fractions of the molecular states \([\Xi_c\bar{D}^\ast]_{1/2}\) and \([\Xi_c\bar{D}^\ast]_{3/2}\) for both Case 1 and Case 2, respectively. It can be seen that the dominant decay channel is \(\Lambda_c\bar{D}_s^\ast\), while the contributions from the \(\Xi_c\bar{D}\) and \(\Lambda_c\bar{D}_s\) channels are negligible in both cases. The main reason is as follows: For the \([\Xi_c\bar{D}^\ast]_{1/2}\) state, its spin wave function can be decomposed under the heavy quark spin symmetry basis as
\begin{eqnarray}
|[\Xi_c\bar{D}^\ast]_{1/2}\rangle=\frac{\sqrt{3}}{2}|0_h\otimes\frac{1}{2}_l\rangle-\frac{1}{2}|1_h\otimes\frac{1}{2}_l\rangle,
\end{eqnarray} 
while those of the \(\Xi_c\bar{D}\) and \(\Lambda_c\bar{D}_s\) channels satisfy
\begin{eqnarray}
|[\Xi_c\bar{D}]_{1/2}\rangle =|[\Lambda_c\bar{D}_s]_{1/2}\rangle = \frac{1}{2}|0_h\otimes\frac{1}{2}_\ell\rangle+\frac{\sqrt{3}}{2}|1_h\otimes\frac{1}{2}_\ell\rangle.\nonumber\\
\end{eqnarray}
It can be observed that the spin wave functions of the initial and final states satisfy a relation analogous to Eq.~\eqref{eq:hqss1}, namely
\begin{eqnarray}\label{eq:hqss2}
\langle [\Xi_c\bar{D}]_{1/2}|\mathcal{L}_{\mathrm{HQSS}}|[\Xi_c\bar{D}^\ast]_{1/2}\rangle&=&0,\\
\langle [\Lambda_c\bar{D}_s]_{1/2}|\mathcal{L}_{\mathrm{HQSS}}|[\Xi_c\bar{D}^\ast]_{1/2}\rangle&=&0.
\end{eqnarray}
For a detailed explanation, see the discussion around Eq.~\eqref{eq:hqss1}.

The reason why the decays of the \( [\Xi_c \bar{D}^\ast]_{3/2} \) state into \( \Xi_c \bar{D} \) and \( \Lambda_c \bar{D}_s \) are suppressed is different from the case of \( [\Xi_c \bar{D}^\ast]_{1/2} \) mentioned above. After all, \( |[\Xi_c\bar{D}^\ast]_{3/2}\rangle=|1_h\otimes\frac{1}{2}_l\rangle \) holds, and there is no orthogonality relation analogous to Eq.~\eqref{eq:hqss2}. The suppression mainly arises from the fact that \( [\Xi_c \bar{D}^\ast]_{3/2} \) couples to \( \Xi_c \bar{D} \) and \( \Lambda_c \bar{D}_s \) via the D-wave, i.e., the suppression is due to higher partial wave contributions.

\begin{table*}
\centering
\renewcommand{\arraystretch}{1.5}
\caption{The ranges of partial decay widths for $[\Xi_c\bar{D}^\ast]_{1/2}$ and $[\Xi_c\bar{D}^\ast]_{3/2}$ in Case 1 [in Eq.~\eqref{eq:PcsCase1}] and Case 2 [in Eq.~\eqref{eq:PcsCase2}] using different form factors (in units of MeV).\label{tab:parPcs4459}}
\setlength{\tabcolsep}{2.85mm}
{
	\begin{tabular}{cccccccc
		}
		\hline\hline
		\multicolumn{1}{c}{} & \multicolumn{1}{c}{Form factors} 	&   With $\mathfrak{F}_1$ &   With $\mathfrak{F}_2$ &   With $\mathfrak{F}_3$ & \multicolumn{2}{c}{Results in other works}  & LHCb~\cite{LHCb:2020jpq}\\
		\hline
		
		\multirow{16}{*}{\textbf{Case 1}} 
		&  $[\Xi_c\bar{D}^\ast]_{1/2}\to\Xi_c\bar{D}$   &  $-$ & $-$ & $-$& $1.9$~\cite{Ortega:2022uyu}& $3.26$~\cite{Shen:2019evi}&   \\ 
		& $[\Xi_c\bar{D}^\ast]_{1/2}\to\Xi^\prime_c\bar{D}$  & $1.7-5.3$ &$1.5-5.6$& $0.6-3.4$& $-$& $-$&  \\  
		& $[\Xi_c\bar{D}^\ast]_{1/2}\to\Lambda_c\bar{D}_s$ &    $-$ & $-$ & $-$& $3.4$~\cite{Ortega:2022uyu}& $3.862$~\cite{Shen:2019evi}&\\ 
		& $[\Xi_c\bar{D}^\ast]_{1/2}\to\Lambda_c\bar{D}^\ast_s$ & $9.3-25.9$ & $7.4-23.04$ & $1.5-7.8$& $16.6$~\cite{Ortega:2022uyu}& $-$&\\
		& $[\Xi_c\bar{D}^\ast]_{1/2}\to \Lambda J/\psi$  & $0.1-0.5$ & $0.1-0.5$ & $0.03-0.2$& $1.6$~\cite{Ortega:2022uyu}&$2.38$~\cite{Shen:2019evi}&   \\
		& $[\Xi_c\bar{D}^\ast]_{1/2}\to\Lambda\eta_c$ &$0.08-0.3$ &$0.07-0.3$& $0.07-0.5$& $0.7$~\cite{Ortega:2022uyu}&$0.076$~\cite{Shen:2019evi}&   \\ 
		
		\cmidrule{2-8} 	&\multicolumn{1}{c}{Total width}  &$11.2-31.9$ &$9.06-29.4$ &$2.2-11.9$& $24.2$~\cite{Ortega:2022uyu} & $42.429$~\cite{Shen:2019evi}& $17.3\pm6.5^{+8}_{-5.7}$  \\	
		\cmidrule{2-8}
		
		&$[\Xi_c\bar{D}^\ast]_{3/2}\to\Xi_c\bar{D}$   &  $-$ & $-$ & $-$& $-$& $0.002$~\cite{Shen:2019evi}&  \\
		& $[\Xi_c\bar{D}^\ast]_{3/2}\to\Xi^\prime_c\bar{D}$  &  $0.1-0.4$ & $0.1-0.3$ & $0.008-0.03$& $-$& $-$&    \\  
		& $[\Xi_c\bar{D}^\ast]_{3/2}\to\Lambda_c\bar{D}_s$&  $-$ & $-$ & $-$& $-$& $0.021$~\cite{Shen:2019evi}&   \\ 
		& $[\Xi_c\bar{D}^\ast]_{3/2}\to\Lambda_c\bar{D}^\ast_s$  & $2.9-26.2$ & $2.3-22.5$ & $0.4-6.1$& $4.8$~\cite{Hu:2021nvs}& $-$&  \\
		& $[\Xi_c\bar{D}^\ast]_{3/2}\to \Lambda J/\psi$ & $0.008-0.07$ & $0.007-0.06$ & $0.001-0.04$&$19.6$~\cite{Hu:2021nvs}& $0.592$~\cite{Shen:2019evi}&  \\
		& $[\Xi_c\bar{D}^\ast]_{3/2}\to\Lambda\eta_c$  &$0.007-0.06$&$0.006-0.06$ &$ 0.001-0.007$&$-$& $0.002$~\cite{Shen:2019evi}&    \\ 
		\cmidrule{2-8}&\multicolumn{1}{c}{Total width}&$3.02-26.8$ &$2.4-23.0$ &$0.4-6.2$&$24.4$~\cite{Hu:2021nvs}  &$3.533$~\cite{Shen:2019evi} & $17.3\pm6.5^{+8}_{-5.7}$ \\
		
		\hline 
		\multirow{16}{*}{\textbf{Case 2}}
		& $[\Xi_c\bar{D}^\ast]_{1/2}\to\Xi_c\bar{D}$   &  $-$ & $-$ & $-$& $1.9$~\cite{Ortega:2022uyu}& $3.26$~\cite{Shen:2019evi}&    \\
		& $[\Xi_c\bar{D}^\ast]_{1/2}\to\Xi^\prime_c\bar{D}$  & $0.5-5.5$ &$0.4-5.4$& $0.2-2.8$& $-$& $-$&  \\
		& $[\Xi_c\bar{D}^\ast]_{1/2}\to\Lambda_c\bar{D}_s$ &    $-$ & $-$ & $-$& $3.4$~\cite{Ortega:2022uyu}& $3.862$~\cite{Shen:2019evi}&  \\
		& $[\Xi_c\bar{D}^\ast]_{1/2}\to\Lambda_c\bar{D}^\ast_s$ & $2.5-26.5$ & $2.3-22.6$ & $0.4-6.2$& $16.6$~\cite{Ortega:2022uyu}& $-$& \\
		& $[\Xi_c\bar{D}^\ast]_{1/2}\to \Lambda J/\psi$  & $0.04-0.5$ & $0.04-0.5$ & $0.008-0.2$& $1.6$~\cite{Ortega:2022uyu}&$2.38$~\cite{Shen:2019evi}& \\
		& $[\Xi_c\bar{D}^\ast]_{1/2}\to\Lambda\eta_c$ &$0.02-0.3$ &$0.02-0.3$& $0.02-0.3$& $0.7$~\cite{Ortega:2022uyu}&$0.076$~\cite{Shen:2019evi}&  \\
		\cmidrule{2-8}
		& \multicolumn{1}{c}{Total width}&$3.4-32.7$ &$2.8-28.7$ &$0.5-9.5$& $24.2$~\cite{Ortega:2022uyu} & $42.429$~\cite{Shen:2019evi} & $17.3\pm6.5^{+8}_{-5.7}$ \\	\cmidrule{2-8}
		\cmidrule{2-8}
		
		& $[\Xi_c\bar{D}^\ast]_{3/2}\to\Xi_c\bar{D}$   &  $-$ & $-$ & $-$& $-$& $0.002$~\cite{Shen:2019evi}&  \\
		& $[\Xi_c\bar{D}^\ast]_{3/2}\to\Xi^\prime_c\bar{D}$  &  $0.1-0.5$ & $0.01-0.4$ & $0.008-0.04$& $-$& $-$&  \\
		& $[\Xi_c\bar{D}^\ast]_{3/2}\to\Lambda_c\bar{D}_s$&  $-$ & $-$ & $-$& $-$& $0.021$~\cite{Shen:2019evi}&  \\
		& $[\Xi_c\bar{D}^\ast]_{3/2}\to\Lambda_c\bar{D}^\ast_s$  & $9.3-26.0$ & $7.4-22.7$ & $1.5-7.7$& $4.8$~\cite{Hu:2021nvs}& $-$& \\
		& $[\Xi_c\bar{D}^\ast]_{3/2}\to \Lambda J/\psi$ & $0.03-0.07$ & $0.02-0.06$ & $0.007-0.05$&$19.6$~\cite{Hu:2021nvs}& $0.592$~\cite{Shen:2019evi}&  \\
		& $[\Xi_c\bar{D}^\ast]_{3/2}\to\Lambda\eta_c$  &$0.02-0.06$&$0.02-0.05$ &$ 0.001-0.008$&$-$& $0.002$~\cite{Shen:2019evi}&  \\
		
		\cmidrule{2-8}
		& \multicolumn{1}{c}{Total width}  &$9.7-26.3$ &$7.7-23.01$ &$1.5-7.8$& $24.4$~\cite{Hu:2021nvs} &$3.533$~\cite{Shen:2019evi} & $17.3\pm6.5^{+8}_{-5.7}$\\
\hline \hline
	\end{tabular}
}
\end{table*}

\begin{table*}[htbp]
\renewcommand{\arraystretch}{1.5}
	\centering
	\caption{The ranges of branching fractions for $[\Xi_c\bar{D}^{(\ast)}]_{1/2}$ and $[\Xi_c\bar{D}^{(\ast)}]_{3/2}$ in Case 1 [in Eq.~\eqref{eq:PcsCase1}] and Case 2 [in Eq.~\eqref{eq:PcsCase2}] using different form factors. \label{tab:fraPcs4459}}
\setlength{\tabcolsep}{3.4mm}
{
	\begin{tabular}{ccccccc}
\hline
\hline
		\multicolumn{1}{c}{} & \multicolumn{3}{c}{\textbf{Case 1}} & \multicolumn{3}{c}{\textbf{Case 2}} \\
		\cmidrule(lr){2-4} \cmidrule(lr){5-7}
		\multicolumn{1}{c}{Form factor} & With $\mathfrak{F}_1$ & With $\mathfrak{F}_2$ & With $\mathfrak{F}_3$ 
		& With $\mathfrak{F}_1$ & With $\mathfrak{F}_2$ & With $\mathfrak{F}_3$ \\
\hline
		$\Gamma_{\Xi_c\bar{D}}/\Gamma_{[\Xi_c\bar{D}^\ast]_{1/2^-}}$ & $-$ &$-$ & $-$ &$-$&$-$&$-$\\ 	$\Gamma_{\Xi^\prime_c\bar{D}}/\Gamma_{[\Xi_c\bar{D}^\ast]_{1/2^-}}$  &  $0.14-0.17$ & $0.15-0.18$ & $0.28-0.3$&$0.14-0.19$&$0.15-0.21$&$0.28-0.31$ \\	$\Gamma_{\Lambda_c\bar{D}_s}/\Gamma_{[\Xi_c\bar{D}^\ast]_{1/2^-}}$  & $-$ & $-$ & $-$&$-$&$-$&$-$ \\$\Gamma_{\Lambda_c\bar{D}^\ast_s}/\Gamma_{[\Xi_c\bar{D}^\ast]_{1/2^-}}$  & $0.81-0.84$ & $0.79-0.83$ &$0.64-0.67$&$0.78-0.84$&$0.77-0.84$&$0.64-0.67$ \\		$\Gamma_{\Lambda J/\psi}/\Gamma_{[\Xi_c\bar{D}^\ast]_{1/2^-}}$  &  $0.01$ &$0.01-0.02$ &$0.02$&$0.007-0.02$&$0.01-0.02$&$0.01-0.02$ \\ 	$\Gamma_{\Lambda \eta_c}/\Gamma_{[\Xi_c\bar{D}^\ast]_{1/2^-}}$   &$0.01$ & $0.01$ & $0.03$&$0.01$&$0.01$&$0.03$ \\	
\hline 
$\Gamma_{\Xi_c\bar{D}}/\Gamma_{[\Xi_c\bar{D}^\ast]_{3/2^-}}$ &  $-$ & $-$&$-$&$-$&$-$&$-$ \\ $\Gamma_{\Xi^\prime_c\bar{D}}/\Gamma_{[\Xi_c\bar{D}^\ast]_{3/2^-}}$  &  $0.02-0.05$ & $0.02-0.05$ & $0.01$&$0.006-0.04$&$0.02-0.05$&$0.007$ \\	 $\Gamma_{\Lambda_c\bar{D}_s}/\Gamma_{[\Xi_c\bar{D}^\ast]_{3/2^-}}$  &  $-$ & $-$ & $-$&$-$&$-$&$-$ \\  $\Gamma_{\Lambda_c\bar{D}^\ast_s}/\Gamma_{[\Xi_c\bar{D}^\ast]_{3/2^-}}$  &  $0.94-0.97$ & $0.95-0.97$ & $0.98$&$0.96-0.99$&$0.95-0.98$&$0.99$\\ $\Gamma_{\Lambda J/\psi}/\Gamma_{[\Xi_c\bar{D}^\ast]_{3/2^-}}$ & $ 0.003$ &$0.003$ & $0.004-0.006$&$0.003$&$0.001-0.003$&$0.005-0.006$ \\ $\Gamma_{\Lambda \eta_c}/\Gamma_{[\Xi_c\bar{D}^\ast]_{3/2^-}}$ &  $0.002$ & $0.002$ & $0.001$&$0.002$&$0.002$&$0.001$ \\
\hline
\hline
	\end{tabular}
}
\end{table*}

\section{Summary}\label{sec:sum}

In recent years, the LHCb Collaboration has reported a series of hidden-charm pentaquark states in the energy region above $4$ GeV, among which the most prominent are \(P_{\psi}^N(4312)\), \(P_{\psi}^N(4440)\) and \(P_{\psi}^N(4457)\), as well as the strange-containing states \(P_{\psi s}^\Lambda(4338)\) and \(P_{\psi s}^\Lambda(4459)\). Their masses lie close to the thresholds of charmed baryons (e.g., \(\Sigma_c\), \(\Xi_c\)) and anti-charmed mesons (e.g., \(\bar{D}\), \(\bar{D}^*\)), and their widths are generally narrow (ranging from a few MeV to about 20 MeV), suggesting that they are very likely molecular states composed of \(\Sigma_c\bar{D}^{(*)}\) and \(\Xi_c\bar{D}^{(*)}\). Within the molecular picture, an unresolved issue concerning the \(P_{\psi}^N\) states is the spin-parity (\(J^P\)) assignment for \(P_{\psi}^N(4440)\) and \(P_{\psi}^N(4457)\), i.e., which one corresponds to \(J^P = 1/2^-\) and which to \(3/2^-\). For \(P_{\psi s}^\Lambda(4459)\), it remains an open question whether the current experimental lineshape contains two substructures with \(J = 1/2\) and \(J = 3/2\).

In this work, we investigate the strong decay behaviors of the \(P_{\psi}^N\) and \(P_{\psi s}^\Lambda\) states within the framework of effective Lagrangians and triangle diagrams. We first construct the effective Lagrangians describing the S-wave couplings between the pentaquark states and their constituent hadrons, where the coupling constants are determined from the residues of the scattering \(T\)-matrix at the bound-state poles. Three commonly used types of form factors—Heaviside, Gaussian, and Multipole—are introduced to regularize the divergences appearing in the loop integrals, and we present in detail the dependence of the results on the cutoff parameters in these form factors. We find that the decay widths increase with the cutoff and are relatively sensitive to it, whereas the branching fractions exhibit a relatively weak dependence. The calculated widths compatible with experimental measurements correspond to relatively small cutoff values, indicating a large interaction range within the system, which is consistent with the characteristics of loosely bound molecular states.

Our results show that \(P_{\psi}^N(4312)\), as an S-wave \(\Sigma_c\bar{D}\) molecular state, has a single dominant decay channel, namely \(\Lambda_c\bar{D}^*\) (with a branching fraction of about $99\%$), while the contribution from the \(\Lambda_c\bar{D}\) channel is suppressed by heavy quark spin symmetry and hence negligible. Given that the experimental width of \(P_{\psi}^N(4312)\) has a small uncertainty and that the state has only one dominant decay channel, the theoretical uncertainties in our calculation are relatively small. Therefore, we use \(P_{\psi}^N(4312)\) as a relatively clean probe to constrain the range of cutoff parameters, and subsequently apply this constrained range to study the decay behaviors of other systems considered in this work. When the spin assignment in Eq.~\eqref{eq:PcCase1} is adopted for \(P_{\psi}^N(4440)\) and \(P_{\psi}^N(4457)\), the width of \(P_{\psi}^N(4457)\) significantly exceeds the experimental upper limit. However, with the assignment in Eq.~\eqref{eq:PcCase1}, the results are consistent with experiment for all three regularization schemes. Hence, our calculation supports the assignment where the lower-mass state \(P_{\psi}^N(4440)\) carries higher spin \(J = 3/2\), and the higher-mass state \(P_{\psi}^N(4457)\) carries lower spin \(J = 1/2\).

Furthermore, we interpret the strange hidden-charm pentaquark \(P_{\psi s}^\Lambda(4338)\) as a \(\Xi_c\bar{D}\) molecular state and find that its width, within the constrained parameter range, is consistent with experimental measurements. Its dominant decay channel is \(\Lambda_c\bar{D}_s\) (with a branching fraction exceeding $90\%$). Taking the experimentally measured width of \(P_{\psi s}^\Lambda(4459)\) as a reference, we also calculate the \(\Xi_c\bar{D}^*\) system for both \(J = 1/2\) and \(J = 3/2\). It is found that the contributions from the \(\Xi_c\bar{D}\) and \(\Lambda_c\bar{D}_s\) channels are suppressed for both molecular states: for \([\Xi_c\bar{D}^*]_{1/2}\), the suppression originates from heavy quark spin symmetry, while for \([\Xi_c\bar{D}^*]_{3/2}\), it is due to D-wave coupling. In addition, the dominant decay channel for both states is \(\Lambda_c\bar{D}_s^*\).

The branching fractions obtained in this work will be helpful for future experimental tests of the molecular interpretation of these pentaquark states, as well as for determining the spins of \(P_{\psi}^N(4440)\) and \(P_{\psi}^N(4457)\). We also hope that, with more accumulated data, a detailed analysis of the lineshape of \(P_{\psi s}^\Lambda(4459)\) can clarify whether it contains two states. Moreover, experimental measurements of the precise contributions from the suppressed channels will also help to assess the role of heavy quark symmetry breaking effects in molecular states.

\section*{Acknowledgement}
This work is supported by the Natural Science Foundation of Hebei Province under Grants No. A2025201016, and the National Natural Science Foundation of China under Grants No. 12235008.

\begin{appendices}
		\begin{widetext}
			\section{decay amplitudes}\label{sec:ap}
The decay amplitudes corresponding to the Feynman diagrams in Fig.~\ref{fig:decays} are given by the following expressions:
			\numberwithin{equation}{section}
			\setcounter{equation}{0}
\begin{eqnarray}
	i\mathcal{M}_{(1-a)}^{[\rho]}&=&iC_{(1-a)}^{[\rho]}\int\frac{d^{4}q}{(2\pi)^{4}}\bar{u}_{\Lambda_c}\frac{\epsilon_{\alpha\beta\gamma\delta}p_1^\alpha q^\beta\gamma^\delta\gamma_5}{\slashed{p}_1+\slashed{q}-m_{\Sigma_c}+i\epsilon}\frac{g^{\gamma\mu}-q^\gamma q^\mu /m_\rho^2}{q^2-m^2_{\rho}+i\epsilon}\frac{(2p_2-q)_\mu}{(p_2-q)^2-m_D^2+i\epsilon}u_{\psi},\\
	i\mathcal{M}_{(1-b)}^{[\pi]}&=&iC_{(1-b)}^{[\pi]}\int\frac{d^{4}q}{(2\pi)^{4}}\bar{u}_{\Lambda_c}\frac{(\gamma^\alpha-(p_1+q)^\alpha/m_{\Sigma_c})\gamma_5 q_\alpha q_\mu\varepsilon_{\bar{D}^\ast}^{\dagger\mu}}{[\slashed{p}_1+\slashed{q}-m_{\Sigma_c}+i\epsilon][q^2-m^2_{\pi}+i\epsilon][(p_2-q)^2-m_D^2+i\epsilon]}u_{\psi},\\ i\mathcal{M}_{(1-b)}^{[\rho]}&=&iC_{(1-b)}^{[\rho]}\int\frac{d^{4}q}{(2\pi)^{4}}\bar{u}_{\Lambda_c}\frac{\epsilon_{\alpha\beta\gamma\delta}p_1^\alpha q^\beta \gamma^\delta\gamma_5}{\slashed{p}_1+\slashed{q}-m_{\Sigma_c}+i\epsilon}\frac{g^{\gamma\nu}-q^\gamma q^\nu /m^2_\rho}{q^2-m^2_{\rho}+i\epsilon}\frac{\epsilon_{\mu\nu\sigma\tau}q^\mu \varepsilon_{\bar{D}^\ast}^{\dagger\sigma}p_2^\tau}{(p_2-q)^2-m_D^2+i\epsilon}u_{\psi},\\
	i\mathcal{M}_{(1-c)}^{[D]}&=&C_{(1-c)}^{[D]}\int\frac{d^{4}q}{(2\pi)^{4}}\bar{u}_{p}\frac{\gamma_5(p_2-2q)_\mu\epsilon_{\psi}^{\dagger\mu}}{\left[\slashed{p}_1+\slashed{q}-m_{\Sigma_c}+i\epsilon\right]\left[q^2-m^2_{D}+i\epsilon\right]\left[(p_2-q)^2-m_{D}^2+i\epsilon\right]}u_{\psi},\\i\mathcal{M}_{(1-c)}^{[D^\ast]}&=&iC_{(1-c)}^{[D^\ast]}\int\frac{d^{4}q}{(2\pi)^{4}}\bar{u}_{p}\frac{\varepsilon_{\alpha\beta\gamma\delta}\gamma_\mu q^\alpha\varepsilon_{\psi}^{\dagger\beta}p_2^\delta(g^{\mu\gamma}-q^\mu q^\gamma/m_{D^\ast}^2)}{\left[\slashed{p}_1+\slashed{q}-m_{\Sigma_c}+i\epsilon\right]\left[q^2-m^2_{D^\ast}+i\epsilon\right]\left[(p_2-q)^2-m_{D}^2+i\epsilon\right]}u_{\psi},\\i\mathcal{M}_{(1-d)}^{[D^\ast]}&=&iC_{(1-d)}^{[D^\ast]}\int\frac{d^{4}q}{(2\pi)^{4}}\bar{u}_{p}\frac{\gamma_\alpha(p_1+q)_\beta\left(g^{\alpha\beta}-q^\alpha q^\beta /m^2_{D^\ast}\right)}{\left[\slashed{p}_1+\slashed{q}-m_{\Sigma_c}+i\epsilon\right]\left[q^2-m^2_{D^\ast}+i\epsilon\right]\left[(p_2-q)^2-m_{D}^2+i\epsilon\right]}u_{\psi},\\i\mathcal{M}_{(2-a)}^{[\pi]}&=&iC_{(2-a)}^{[\pi]}\int\frac{d^{4}q}{(2\pi)^{4}}\bar{u}_{\Lambda_c}\frac{g^{\beta\omega}-(p_2-q)^\beta(p_2-q)^\omega/m_{D^\ast}^2}{[\slashed{p}_1+\slashed{q}-m_{\Sigma_c}+i\epsilon][q^2-m^2_{\pi}+i\epsilon][(p_2-q)^2-m_{D^\ast}^2+i\epsilon]}\nonumber\\ &&\times (\gamma^\alpha-(p_1+q)^\alpha/m_{\Sigma_c})\gamma_5 q_{\alpha} q_\beta \mathcal{U}_\omega^J,\\ i\mathcal{M}_{(2-a)}^{[\rho]}&=&iC_{(2-a)}^{[\rho]}\int\frac{d^{4}q}{(2\pi)^{4}}\bar{u}_{\Lambda_c}\frac{[g^{\gamma\nu}-q^\gamma q^\nu /m^2_\rho][g^{\sigma\omega}-(p_2-q)^\sigma(p_2-q)^\omega/m_{D^\ast}]}{[\slashed{p}_1+\slashed{q}-m_{\Sigma_c}+i\epsilon][q^2-m^2_{\pi}+i\epsilon][(p_2-q)^2-m_{D^\ast}^2+i\epsilon]}\nonumber\\&&\times \epsilon_{\alpha\beta\gamma\delta} \epsilon_{\mu\nu\sigma\tau}p_1^\alpha q^\beta\gamma^\delta\gamma_5 q^\mu p_2^\tau \mathcal{U}_\omega^J,\\
	i\mathcal{M}_{(2-b)}^{[\pi]}&=&iC_{(2-b)}^{[\pi]}\int\frac{d^{4}q}{(2\pi)^{4}}\bar{u}_{\Lambda_c}\frac{g^{\omega\nu}-(p_2-q)^\omega(p_2-q)^\nu/m_{D^\ast}^2}{[\slashed{p}_1+\slashed{q}-m_{\Sigma_c}+i\epsilon][q^2-m^2_{\pi}+i\epsilon][(p_2-q)^2-m_{D^\ast}^2+i\epsilon]}\nonumber\\ &&\times \epsilon_{\mu\nu\sigma\tau}q^\mu\varepsilon^{\dagger\sigma}_{\bar{D}^{\ast}}p_2^\tau(\gamma^\alpha-(p_1+q)^\alpha/m_{\Sigma_c})\gamma_5 q_{\alpha} \mathcal{U}_\omega^J,\\i\mathcal{M}_{(2-b)}^{[\rho]}&=&iC_{(2-b)}^{[\rho]}\int\frac{d^{4}q}{(2\pi)^{4}}\bar{u}_{\Lambda_c}\frac{\epsilon_{\alpha\beta\gamma\delta}p_1^\alpha q^\beta\gamma^\delta\gamma_5}{\left[\slashed{p}_1+\slashed{q}-m_{\Sigma_c}+i\epsilon\right]\left[q^2-m^2_{\rho}+i\epsilon\right]\left[(p_2-q)^2-m_{D^\ast}^2+i\epsilon\right]}\nonumber\\ &&\times\sum_{\lambda_{1},\lambda_{2}}\varepsilon_{\rho}^{\dagger\gamma}(\lambda_{1})\epsilon_{\bar{D}^\ast}^{\dagger\omega}(\lambda_2)[-\frac{1}{2}\beta(2p_2-q)_\nu\varepsilon_\rho^\nu(\lambda_1)\varepsilon_{\bar{D}^{\ast}\mu}(\lambda_2) \varepsilon_{\bar{D}^\ast}^{\dagger\mu} \nonumber\\ && +2m_{D^\ast}\lambda\varepsilon_{\bar{D}^{\ast}}^\mu (\lambda_2)\varepsilon^{\dagger\nu}_{\bar{D}^\ast}(q_\mu\varepsilon_{1\nu}(\lambda_1)-q_\nu\varepsilon_{1\mu}(\lambda_1)]\mathcal{U}_\omega^J,\\i\mathcal{M}_{(2-c)}^{[\pi]}&=&C_{(2-c)}^{[\pi]}\int\frac{d^{4}q}{(2\pi)^{4}}\bar{u}_{\Sigma_c}\frac{\epsilon_{\mu\nu\alpha\beta}\gamma^\mu q^\nu\gamma^\alpha p_1^\beta q_\sigma (g^{\sigma\omega}-(p_2-q)^\sigma(p_2-q)^\omega/m_{D^\ast}^2)}{\left[\slashed{p}_1+\slashed{q}-m_{\Sigma_c}+i\epsilon\right]\left[q^2-m^2_{\rho}+i\epsilon\right]\left[(p_2-q)^2-m_{D^\ast}^2+i\epsilon\right]}\mathcal{U}_\omega^J,\\i\mathcal{M}_{(2-c)}^{[\rho]}&=&C_{(2-c)}^{[\rho]}\int\frac{d^{4}q}{(2\pi)^{4}}\bar{u}_{\Sigma_c}\frac{\epsilon_{\alpha\beta\gamma\delta}q^\alpha p_2^\delta(g^{\gamma\omega}-(p_2-q)^\gamma(p_2-q)^\omega/m_{D^\ast}^2)}{\left[\slashed{p}_1+\slashed{q}-m_{\Sigma_c}+i\epsilon\right]\left[q^2-m^2_{\rho}+i\epsilon\right]\left[(p_2-q)^2-m_{D^\ast}^2+i\epsilon\right]}\nonumber\\ &&\times\sum_{\lambda_{1}}\varepsilon_{1\beta}(\lambda_{1})[\lambda_s(\gamma_\mu-p_{1\mu}/m_{\Sigma_c})(q^\mu\varepsilon_{1\nu}^{\dagger}(\lambda_{1})-q_\nu\varepsilon_1^{\dagger\mu}(\lambda_{1}))(\gamma_\nu+(p_1+q)_{\nu}/m_{\Sigma_c})\nonumber\\ &&-\frac{1}{2}\beta_s(5-2\gamma_{\mu}p_1^{\mu}/m_{\Sigma_c})(2p_1+q)_\nu/m_{\Sigma_c}\varepsilon_1^{\dagger\nu}(\lambda_{1})]\mathcal{U}_\omega^J,\\i\mathcal{M}_{(2-d)}^{[\pi]}&=&C_{(2-d)}^{[\pi]}\int\frac{d^{4}q}{(2\pi)^{4}}\bar{u}^\mu_{\Sigma_c^\ast}\frac{\epsilon_{\mu\nu\alpha\beta} q^\nu\gamma^\alpha p_1^\beta q_\sigma (g^{\sigma\omega}-(p_2-q)^\sigma(p_2-q)^\omega/m_{D^\ast}^2)}{\left[\slashed{p}_1+\slashed{q}-m_{\Sigma_c}+i\epsilon\right]\left[q^2-m^2_{\pi}+i\epsilon\right]\left[(p_2-q)^2-m_{D^\ast}^2+i\epsilon\right]}\mathcal{U}_\omega^J,\\i\mathcal{M}_{(2-d)}^{[\rho]}&=&C_{(2-d)}^{[\rho]}\int\frac{d^{4}q}{(2\pi)^{4}}\bar{u}_{\Sigma_c^\ast}^\mu\frac{\epsilon_{\alpha\beta\gamma\delta}q^\alpha p_2^\delta(g^{\gamma\omega}-(p_2-q)^\gamma(p_2-q)^\omega/m_{D^\ast}^2)}{\left[\slashed{p}_1+\slashed{q}-m_{\Sigma_c}+i\epsilon\right]\left[q^2-m^2_{\rho}+i\epsilon\right]\left[(p_2-q)^2-m_{D^\ast}^2+i\epsilon\right]}\nonumber\\ &&\times\sum_{\lambda_{1}}\varepsilon_{1}^\beta(\lambda_{1})[\lambda_s(q_\mu\varepsilon_{1}^{\dagger^\nu}(\lambda_{1})-q^\nu\varepsilon_{1\mu}^{\dagger}(\lambda_{1})) (\gamma_\nu-(p_1-q)_\nu/m_{\Sigma_c})\nonumber\\ &&+\beta_s(\gamma_\mu- (p_1+q)_\mu/m_{\Sigma_c})\gamma_5 (p_1+q)_\nu/m_{\Sigma_c}\varepsilon_\rho^{\dagger\nu} ]\mathcal{U}_\omega^J,\\i\mathcal{M}_{(2-e)}^{[D]}&=&C_{(2-e)}^{[D]}\int\frac{d^{4}q}{(2\pi)^{4}}\bar{u}_{p}\frac{\epsilon_{\alpha\beta\gamma\delta}\gamma_5 q^\alpha \varepsilon^\beta_{\psi}p_2^\delta(g^{\gamma\omega}-(p_2-q)^\gamma(p_2-q)^\omega /m_{D^\ast})}{\left[\slashed{p}_1+\slashed{q}-m_{\Sigma_c}+i\epsilon\right]\left[q^2-m^2_{D}+i\epsilon\right]\left[(p_2-q)^2-m_{D^\ast}^2+i\epsilon\right]}\mathcal{U}_\omega^J \\i\mathcal{M}_{(2-e)}^{[D^\ast]}&=&iC_{(2-e)}^{[D^\ast]}\int\frac{d^{4}q}{(2\pi)^{4}}\bar{u}_{p}\frac{\gamma_\alpha(p_2-2q)_\mu}{\left[\slashed{p}_1+\slashed{q}-m_{\Sigma_c}+i\epsilon\right]\left[q^2-m^2_{D^\ast}+i\epsilon\right]\left[(p_2-q)^2-m_{D^\ast}^2+i\epsilon\right]}\nonumber\\ &&\times\sum_{\lambda_{2},\lambda_{3}}\varepsilon_{2}^{\dagger\alpha}(\lambda_{2})\varepsilon_{3}^{\dagger\omega}(\lambda_{3})\left[(\varepsilon_{\psi }^{\dagger\nu}\varepsilon_{2\nu}(\lambda_{2})\varepsilon_{3}^\mu(\lambda_{3})+\varepsilon_{\psi }^{\dagger\nu}\varepsilon_{2}^\mu(\lambda_{2})\varepsilon_{3}(\lambda_{3})-\varepsilon_{\psi }^{\dagger\mu}\varepsilon_{2\nu}(\lambda_{2})\varepsilon_{3}^\nu(\lambda_{3}))\right]\mathcal{U}_\omega^J,\\i\mathcal{M}_{(2-f)}^{[D]}&=&C_{(2-f)}^{[D]}\int\frac{d^{4}q}{(2\pi)^{4}}\bar{u}_{p}\frac{\gamma_5(p_2-2q)_\nu(g^{\omega\nu}-(p_2-q)^\omega(p_2-q)^\nu)/m_{D^\ast}^2}{\left[\slashed{p}_1+\slashed{q}-m_{\Sigma_c}+i\epsilon\right]\left[q^2-m^2_{D}+i\epsilon\right]\left[(p_2-q)^2-m_{D^\ast}^2+i\epsilon\right]}\mathcal{U}_\omega^J, \\i\mathcal{M}_{(2-f)}^{[D^\ast]}&=&iC_{(2-f)}^{[D^\ast]}\int\frac{d^{4}q}{(2\pi)^{4}}\bar{u}_{p}\frac{[g^{\mu\beta }-q^\mu q^\beta/m_{D^\ast}^2][g^{\gamma\omega}-(p_2-q)^\gamma(p_2-q)^\omega/m_{D^\ast}^2]}{[\slashed{p}_1+\slashed{q}-m_{\Sigma_c}+i\epsilon][q^2-m_{D^\ast}^2+i\epsilon][(p_2-q)^2-m_{D^\ast}^2+i\epsilon]}\nonumber\\ &&\times\epsilon_{\alpha\beta\gamma\delta}\gamma_\mu q^\alpha p_2^\delta\mathcal{U}_\omega^J,\\
	i\mathcal{M}_{(3-a)}^{[\bar{K}^\ast]}&=&C_{(3-a)}^{[\bar{K}^\ast]}\int\frac{d^{4}q}{(2\pi)^{4}}\bar{u}_{\Lambda_c}\frac{(p_1-q)_\alpha\left(g^{\alpha\beta}-q^\alpha q^\beta /m^2_{K^\ast}\right)(2p_2-q)_\beta}{\left[\slashed{p}_1+\slashed{q}-m_{\Xi_c}+i\epsilon\right]\left[q^2-m^2_{K^\ast}+i\epsilon\right]\left[(p_2-q)^2-m_{D}^2+i\epsilon\right]}u_{\psi s},\\	i\mathcal{M}_{(3-b)}^{[D]}&=&C_{(3-b)}^{[D]}\int\frac{d^{4}q}{(2\pi)^{4}}\bar{u}_{\Lambda}\frac{\gamma_5(p_2-2q)_\mu\varepsilon_{\psi}^{\dagger\mu}}{\left[\slashed{p}_1+\slashed{q}-m_{\Xi_c}+i\epsilon\right]\left[q^2-m^2_{D}+i\epsilon\right]\left[(p_2-q)^2-m_{D}^2+i\epsilon\right]}u_{\psi s},\\i\mathcal{M}_{(3-b)}^{[D^\ast]}&=&iC_{(3-b)}^{[D^\ast]}\int\frac{d^{4}q}{(2\pi)^{4}}\bar{u}_{\Lambda}\frac{\epsilon_{\alpha\beta\gamma\delta}\gamma_\mu q^\alpha\varepsilon_{\psi}^{\dagger\beta}p_2^\delta(g^{\mu\gamma}-q^\mu q^\gamma/m_{D^\ast}^2)}{\left[\slashed{p}_1+\slashed{q}-m_{\Xi_c}+i\epsilon\right]\left[q^2-m^2_{D^\ast}+i\epsilon\right]\left[(p_2-q)^2-m_{D}^2+i\epsilon\right]}u_{\psi s},\\i\mathcal{M}_{(3-c)}^{[D^\ast]}&=&iC_{(3-c)}^{[D^\ast]}\int\frac{d^{4}q}{(2\pi)^{4}}\bar{u}_{\Lambda}\frac{\gamma_\alpha(p_2-2q)_\beta\left(g^{\alpha\beta}-q^\alpha q^\beta /m^2_{D^\ast}\right)}{\left[\slashed{p}_1+\slashed{q}-m_{\Xi_c}+i\epsilon\right]\left[q^2-m^2_{D^\ast}+i\epsilon\right]\left[(p_2-q)^2-m_{D}^2+i\epsilon\right]}u_{\psi s},\\i\mathcal{M}_{(4-a)}^{[\rho]}&=&iC_{4-a}^{[\rho]}\int\frac{d^{4}q}{(2\pi)^{4}}\bar{u}_{\Xi_c} \frac{[g^{\alpha\delta}-q^\alpha q^\delta /m_{\rho}^2][g^{\theta\omega }-(p_2-q)^\theta (p_2-q)^\omega/m_{D^\ast}^2] }{[\slashed{p}_1+\slashed{q}-m_{\Xi_c}+i\epsilon][q^2-m_{\rho}^2+i\epsilon][(p_2-q)^2-m_{D^\ast}^2+i\epsilon]} \nonumber\\ &&\times\epsilon_{\gamma\delta\theta\kappa}(p_1-q)_\alpha q^\gamma p_2^\kappa\mathcal{U}_\omega^J ,\\i\mathcal{M}_{(4-b)}^{[\pi]}&=&iC_{(4-b)}^{[\pi]}\int\frac{d^{4}q}{(2\pi)^{4}}\bar{u}_{\Xi_c^\prime}\frac{g^{\beta\omega}-(p_2-q)^\beta(p_2-q)^\omega/m_{D^\ast}^2}{[\slashed{p}_1+\slashed{q}-m_{\Xi_c}+i\epsilon][q^2-m^2_{\pi}+i\epsilon][(p_2-q)^2-m_{D^\ast}^2+i\epsilon]}\nonumber\\ &&\times(\gamma^\alpha-(p_1+q)^\alpha/m_{\Sigma_c})\gamma_5 q_{\alpha}q_\beta\mathcal{U}_\omega^J ,\\ i\mathcal{M}_{(4-b)}^{[\rho]}&=&iC_{(4-b)}^{[\rho]}\int\frac{d^{4}q}{(2\pi)^{4}}\bar{u}_{\Xi_c^\prime}\frac{[g^{\sigma\omega}-(p_2-q)^\sigma(p_2-q)^\omega/m_{D^\ast}][g^{\gamma\nu}-q^\gamma q^\nu /m^2_\rho]}{[\slashed{p}_1+\slashed{q}-m_{\Xi_c}+i\epsilon][q^2-m^2_{\rho}+i\epsilon][(p_2-q)^2-m_{D^\ast}^2+i\epsilon]}\nonumber\\&&\times\epsilon_{\alpha\beta\gamma\delta}\epsilon_{\mu\nu\sigma\tau}p_1^\alpha q^\beta\gamma^\delta\gamma_5 q^\mu p_2^\tau \mathcal{U}_\omega^J,
	\\i\mathcal{M}_{(4-c)}^{[\bar{K}^\ast]}&=&C_{4-c}^{[\bar{K}^\ast]}\int\frac{d^{4}q}{(2\pi)^{4}}\bar{u}_{\Lambda_c}\frac{[g^{\alpha\delta}-q^\alpha q^\delta /m_{\bar{K}^\ast}^2][g^{\theta\omega }-(p_2-q)^\theta (p_2-q)^\omega/m_{D^\ast}^2]}{[\slashed{p}_1+\slashed{q}-m_{\Xi_c}+i\epsilon][q^2-m_{K^\ast}^2+i\epsilon][(p_2-q)^2-m_{D^\ast}^2+i\epsilon]} \nonumber\\ &&\times\epsilon_{\gamma\delta\theta\kappa}(p_1-q)_\alpha q^\gamma p_2^\kappa\mathcal{U}_\omega^J ,\\i\mathcal{M}_{(4-d)}^{[\bar{K}^\ast]}&=&C_{(4-d)}^{[\bar{K}^\ast]}\int\frac{d^{4}q}{(2\pi)^{4}}\bar{u}_{\Lambda_c}\frac{(p_1-q)^\alpha}{\left[\slashed{p}_1+\slashed{q}-m_{\Xi_c}+i\epsilon\right]\left[q^2-m^2_{K^\ast}+i\epsilon\right]\left[(p_2-q)^2-m_{D^\ast}^2+i\epsilon\right]}\nonumber\\ &&\times\sum_{\lambda_{3},\lambda_{4}}\varepsilon_{4\alpha}^{\dagger}(\lambda_{4})\varepsilon_{3}^{\dagger\omega}(\lambda_{3})[4\lambda\sqrt{m_{D^{*}}m_{D_s^{*}}} \varepsilon_{3\beta}(\lambda_3)\varepsilon_{\bar{D}_s^\ast}^{\dagger\gamma}(q^\beta\varepsilon_{4\gamma}(\lambda_4)-q_\gamma\varepsilon_{4}^\beta(\lambda_4))\nonumber\\ &&+\beta\varepsilon_{3\beta}(\lambda_3)\varepsilon_{\bar{D}_s^\ast}^{\dagger\beta}(2p_2-q)^\gamma\varepsilon_{4\gamma}(\lambda_4)]\mathcal{U}_\omega^J ,\\i\mathcal{M}_{(4-e)}^{[D]}&=&C_{(4-e)}^{[D]}\int\frac{d^{4}q}{(2\pi)^{4}}\bar{u}_{\Lambda}\frac{\epsilon_{\alpha\beta\gamma\delta}\gamma_5 q^\alpha \varepsilon^\beta_{\psi}p_2^\delta(g^{\gamma\omega}-(p_2-q)^\gamma(p_2-q)^\omega /m_{D^\ast})}{\left[\slashed{p}_1+\slashed{q}-m_{\Xi_c}+i\epsilon\right]\left[q^2-m^2_{D}+i\epsilon\right]\left[(p_2-q)^2-m_{D^\ast}^2+i\epsilon\right]}\mathcal{U}_\omega^J, \\i\mathcal{M}_{(4-e)}^{[D^\ast]}&=&iC_{(4-e)}^{[D^\ast]}\int\frac{d^{4}q}{(2\pi)^{4}}\bar{u}_{\Lambda}\frac{\gamma_\alpha(p_2-2q)_\mu}{\left[\slashed{p}_1+\slashed{q}-m_{\Xi_c}+i\epsilon\right]\left[q^2-m^2_{D^\ast}+i\epsilon\right]\left[(p_2-q)^2-m_{D^\ast}^2+i\epsilon\right]}\nonumber\\ &&\times\sum_{\lambda_{2},\lambda_{3}}\varepsilon_{2}^{\dagger\alpha}(\lambda_{2})\varepsilon_{3}^{\dagger\omega}(\lambda_{3})\left[(\varepsilon_{\psi }^{\dagger\nu}\varepsilon_{2\nu}(\lambda_{2})\varepsilon_{3}^\mu(\lambda_{3})+\varepsilon_{\psi }^{\dagger\nu}\varepsilon_{2}^\mu(\lambda_{2})\varepsilon_{3\nu}(\lambda_{3})-\varepsilon_{\psi }^{\dagger\mu}\varepsilon_{2\nu}(\lambda_{2})\varepsilon_{3}^\nu(\lambda_{3}))\right]\mathcal{U}_\omega^J,\\i\mathcal{M}_{(4-f)}^{[D]}&=&C_{(4-f)}^{[D]}\int\frac{d^{4}q}{(2\pi)^{4}}\bar{u}_{\Lambda}\frac{\gamma_5(p_2-2q)_\nu(g^{\omega\nu}-(p_2-q)^\omega(p_2-q)^\nu)/m_{D^\ast}^2}{\left[\slashed{p}_1+\slashed{q}-m_{\Xi_c}+i\epsilon\right]\left[q^2-m^2_{D}+i\epsilon\right]\left[(p_2-q)^2-m_{D^\ast}^2+i\epsilon\right]}\mathcal{U}_\omega^J,\\i\mathcal{M}_{(4-f)}^{[D^\ast]}&=&iC_{(4-f)}^{[D^\ast]}\int\frac{d^{4}q}{(2\pi)^{4}}\bar{u}_{\Lambda}\frac{\epsilon_{\alpha\beta\gamma\delta}\gamma_\mu q^\alpha p_2\delta}{\slashed{p}_1+\slashed{q}-m_{\Xi_c}+i\epsilon}\frac{g^{\mu\beta }-q^\mu q^\beta/m_{D^\ast}^2}{q^2-m_{D^\ast}^2+i\epsilon} \frac{g^{\gamma\omega}-(p_2-q)^\gamma(p_2-q)^\omega)/m_{D^\ast}^2 }{(p_2-q)^2-m_{D^\ast}^2+i\epsilon}\mathcal{U}_\omega^J.\label{eq:a33}
\end{eqnarray}			
In the notation such as $\mathcal{M}^{[\pi]}_{(1-a)}$, the subscript indicates the label of the corresponding diagram in Fig.~\ref{fig:decays}, while the superscript denotes that the exchanged particle is $\pi$ meson. The symbols $\varepsilon_\psi$, $\varepsilon_{\bar{D}^\ast}$, and $\varepsilon_{\bar{D}_s^\ast}$, represent the polarization vectors of $J/\psi$, $\bar{D}^\ast$, and $\bar{D}^\ast_s$, respectively. The corresponding coupling constants from the three vertices in Fig.~\ref{fig:decays} are absorbed into coefficients such as $C^{[\pi]}_{(1-a)}$. 
The spinors \(u_{\psi}\) and \(u_{\psi s}\) correspond to the pentaquarks \([\Sigma_c \bar{D}]_{1/2}\) [i.e., \(P_{\psi}^N(4312)\)] and \([\Xi_c \bar{D}]_{1/2}\) [i.e., \(P_{\psi s}^\Lambda(4338)\)], respectively. For the \([\Sigma_c \bar{D}^*]_{1/2,3/2}\) and \([\Xi_c \bar{D}^*]_{1/2,3/2}\) states, the structure of the amplitudes is similar. Therefore, we employ the notation \(\mathcal{U}_\omega^J\), where the subscript \(\omega\) denotes a Lorentz index, and the superscript \(J = 1/2\) or \(3/2\) distinguishes between the pentaquarks with spin \(1/2\) and \(3/2\), respectively. Their expressions are given in the following,
\begin{align}\label{eq:coeffs}
	C_{(1-a)}^{[\pi]}&=-\frac{\sqrt{6}}{3\sqrt{ m_{\Sigma_c}m_{\Lambda_c}}}\lambda_I\beta g_V^2 g_{\psi},&
	C_{(1-b)}^{[\pi]}&=-\frac{\sqrt{6m_{D^\ast }m_D}}{3f_\pi^2}g_4 g_b g_{\psi} ,\nonumber\\ 
	C_{(1-b)}^{[\rho]}&=-4\sqrt{\frac{2}{3m_{\Sigma_c}m_{\Lambda_c}}}\lambda_I \lambda g_V^2 g_{\psi},&	
	C_{(1-c)}^{[D]}&=-4\sqrt{m_D^2 m_{J/\psi}}g_{\Sigma_c DN}g_2g_{\psi},\nonumber\\ 
	C_{(1-c)}^{[D^\ast]}&=\frac{8\sqrt{m_D m_{D^\ast}m_{J/\psi}}}{m_{J/\psi}}g_{ \Sigma_c D^\ast N}g_2g_{\psi},&
	C_{(1-d)}^{[D^\ast]}&=4\sqrt{m_D m_{D^\ast}m_{\eta_c}}g_{ \Sigma_c D^\ast N}g_2g_{\psi},\nonumber\\ 
	C_{(3-a)}^{[\bar{K}^\ast]}&=-\sqrt{\frac{1}{4m_{\Xi_c}m_D}}\beta_B\beta g_V^2g_{\psi s},&
	C_{(3-b)}^{[D]}&=\sqrt{\frac{2m_D^3 m_{J/\psi}}{3}} g_{B_8 B_{\bar{3}}D}g_2g_{\psi s},\nonumber\\
	C_{(3-b)}^{[D^\ast]}&=2\sqrt{\frac{2m_Dm_{D^\ast}^2}{3m_{J/\psi}}} g_{B_8 B_{\bar{3}}D^\ast}g_2g_{\psi s},&
	C_{(3-c)}^{[D^\ast]}&=-\sqrt{\frac{2m_Dm_{D^\ast}^2 m_{\eta_c}}{3}} g_{B_8 B_{\bar{3}}D^\ast}g_2g_{\psi s} ,\nonumber\\ 
	C_{(2-a)}^{[\pi]}\mathcal{U}_{\omega}^{1/2}&=-\frac{\sqrt{6m_D m_{D^\ast}}}{3f_\pi^2}g_4 g_b \tilde{g}_{\psi}\gamma_\omega\gamma_5u_{\psi},&
	C_{(2-a)}^{[\pi]}\mathcal{U}_{\omega}^{3/2}&=-\frac{\sqrt{6m_D m_{D^\ast}}}{3f_\pi^2}g_4 g_b\tilde{g}_{\psi}^{\prime}u_{\psi \omega},\nonumber\\ 	C_{(2-a)}^{[\rho]}\mathcal{U}_{\omega}^{1/2}&=\frac{4\sqrt{6}}{3\sqrt{m_{\Sigma_c}m_{\Lambda_c}}}\lambda_I \lambda g_V^2\tilde{g}_{\psi}\gamma_\omega\gamma_5u_{\psi},&
	C_{(2-a)}^{[\rho]}\mathcal{U}_{\omega}^{3/2}&= \frac{4\sqrt{6}}{3\sqrt{m_{\Sigma_c}m_{\Lambda_c}}}\lambda_I \lambda g_V^2\tilde{g}_{\psi}^{\prime}u_{\psi \omega},\nonumber\\ 	C_{(2-b)}^{[\pi]}\mathcal{U}_{\omega}^{1/2}&=-\frac{\sqrt{6}}{3f_\pi^2}g_4 g_b\tilde{g}_{\psi}\gamma_\omega\gamma_5u_{\psi},&
	C_{(2-b)}^{[\pi]}\mathcal{U}_{\omega}^{3/2}&= -\frac{\sqrt{6}}{3f_\pi^2}g_4 g_b\tilde{g}_{\psi}^{\prime}u_{\psi \omega},\nonumber\\ C_{(2-b)}^{[\rho]}\mathcal{U}_{\omega}^{1/2}&=-\frac{2\sqrt{6}}{3\sqrt{m_{\Sigma_c}m_{\Lambda_c}}}\lambda_I \beta g_V^2\tilde{g}_{\psi}\gamma_\omega\gamma_5u_{\psi},&
	C_{(2-b)}^{[\rho]}\mathcal{U}_{\omega}^{3/2}&=-\frac{2\sqrt{6}}{3\sqrt{m_{\Sigma_c}m_{\Lambda_c}}}\lambda_I \beta g_V^2\tilde{g}_{\psi}^{\prime}u_{\psi \omega},\nonumber\\ C_{(2-c)}^{[\pi]}\mathcal{U}_{\omega}^{1/2}&=\frac{\sqrt{m_D m_{D^\ast}}}{2f_\pi^2 m_{\Sigma_c}}g_1 g_b\tilde{g}_{\psi}\gamma_\omega\gamma_5u_{\psi},&
	C_{(2-c)}^{[\pi]}\mathcal{U}_{\omega}^{3/2}&= \frac{\sqrt{m_D m_{D^\ast}}}{2f_\pi^2 m_{\Sigma_c}}g_1 g_b\tilde{g}_{\psi}^{\prime}u_{\psi \omega},\nonumber\\ C_{(2-c)}^{[\rho]}\mathcal{U}_{\omega}^{1/2}&=\frac{2}{3}\lambda g_V^2\tilde{g}_{\psi}\gamma_\omega\gamma_5u_{\psi},&
	C_{(2-c)}^{[\rho]}\mathcal{U}_{\omega}^{3/2}&= \frac{2}{3}\lambda g_V^2\tilde{g}_{\psi}^{\prime}u_{\psi \omega},\nonumber\\ C_{(2-d)}^{[\pi]}\mathcal{U}_{\omega}^{1/2}&=\frac{\sqrt{3m_D m_{D^\ast}}}{2f_\pi^2 m_{\Sigma_c}}g_1 g_b\tilde{g}_{\psi}\gamma_\omega\gamma_5u_{\psi},&
	C_{(2-d)}^{[\pi]}\mathcal{U}_{\omega}^{3/2}&= \frac{\sqrt{3m_D m_{D^\ast}}}{2f_\pi^2 m_{\Sigma_c}}g_1 g_b\tilde{g}_{\psi}^{\prime}u_{\psi \omega},\nonumber\\ C_{(2-d)}^{[\rho]}\mathcal{U}_{\omega}^{1/2}&=\frac{2\sqrt{3}}{3}\lambda g_V^2\tilde{g}_{\psi}\gamma_\omega\gamma_5u_{\psi},&
	C_{(2-d)}^{[\rho]}\mathcal{U}_{\omega}^{3/2}&=\frac{2\sqrt{3}}{3}\lambda g_V^2\tilde{g}_{\psi}^{\prime}u_{\psi \omega},\nonumber\\ C_{(2-e)}^{[D]}\mathcal{U}_{\omega}^{1/2}&=\frac{8\sqrt{m_D m_{D^\ast}m_{J/\psi}}}{m_{J/\psi}}g_{\Sigma_c DN}g_2\tilde{g}_{\psi}\gamma_\omega\gamma_5u_{\psi},&
	C_{(2-e)}^{[D]}\mathcal{U}_{\omega}^{3/2}&= \frac{8\sqrt{m_D m_{D^\ast}m_{J/\psi}}}{m_{J/\psi}}g_{\Sigma_c DN}g_2\tilde{g}_{\psi}^{\prime}u_{\psi \omega},\nonumber\\ C_{(2-e)}^{[D^\ast]}\mathcal{U}_{\omega}^{1/2}&=4m_{D^\ast}\sqrt{m_{J/\psi}}g_{\Sigma_c D^\ast N }g_2\tilde{g}_{\psi}\gamma_\omega\gamma_5u_{\psi},&
	C_{(2-e)}^{[D^\ast]}\mathcal{U}_{\omega}^{3/2}&=4m_{D^\ast}\sqrt{m_{J/\psi}}g_{\Sigma_c D^\ast N }g_2\tilde{g}_{\psi}^{\prime}u_{\psi \omega},\nonumber\\ C_{(2-f)}^{[D]}\mathcal{U}_{\omega}^{1/2}&=-4\sqrt{m_D m_{D^\ast}m_{\eta_c}}g_{\Sigma_c DN}g_2\tilde{g}_{\psi}\gamma_\omega\gamma_5u_{\psi},&
	C_{(2-f)}^{[D]}\mathcal{U}_{\omega}^{3/2}&=-4\sqrt{m_D m_{D^\ast}m_{\eta_c}}g_{\Sigma_c DN}g_2\tilde{g}_{\psi}^{\prime}u_{\psi \omega},\nonumber\\  C_{(2-f)}^{[D^\ast]}\mathcal{U}_{\omega}^{1/2}&=-\frac{8m_{D^\ast}}{\sqrt{m_{\eta_c}}}g_{\Sigma_c D^\ast N }g_2\tilde{g}_{\psi}\gamma_\omega\gamma_5u_{\psi},&
	C_{(2-f)}^{[D^\ast]}\mathcal{U}_{\omega}^{3/2}&=-\frac{8m_{D^\ast}}{\sqrt{m_{\eta_c}}}g_{\Sigma_c D^\ast N }g_2\tilde{g}_{\psi}^{\prime}u_{\psi \omega},\nonumber\\
	C_{(4-a)}^{[\rho]}\mathcal{U}_{\omega}^{1/2}&=-\sqrt{\frac{1}{2 m_{\Xi_c}^2}}\beta_B\lambda g_V^2\tilde{g}_{\psi s}\gamma_\omega\gamma_5 u_{\psi s},&
	C_{(4-a)}^{[\rho]}\mathcal{U}_{\omega}^{3/2}&=-\sqrt{\frac{1}{2 m_{\Xi_c}^2}}\beta_B\lambda g_V^2\tilde{g}_{\psi s}^{\prime}u_{\psi s  \omega},\nonumber\\  
	C_{(4-b)}^{[\pi]}\mathcal{U}_{\omega}^{1/2}&=-\frac{\sqrt{6m_{D^\ast}m_{D}}}{6f_\pi^2}g_4g_b\tilde{g}_{\psi s}\gamma_\omega\gamma_5 u_{\psi s},&
	C_{(4-b)}^{[\pi]}\mathcal{U}_{\omega}^{3/2}&=-\frac{\sqrt{6m_{D^\ast}m_{D}}}{6f_\pi^2}g_4g_b\tilde{g}_{\psi s}^{\prime}u_{\psi s  \omega},\nonumber\\  
	C_{(4-b)}^{[\rho]}\mathcal{U}_{\omega}^{1/2}&=2\sqrt{\frac{2}{3m_{\Xi_c}m_{\Xi_c^{'}}}}\lambda_I \beta g_V^2\tilde{g}_{\psi s}\gamma_\omega\gamma_5 u_{\psi s},&
	C_{(4-b)}^{[\rho]}\mathcal{U}_{\omega}^{3/2}&=2\sqrt{\frac{2}{3m_{\Xi_c}m_{\Xi_c^{'}}}}\lambda_I \beta g_V^2\tilde{g}_{\psi s}^{\prime}u_{\psi s  \omega},\nonumber\\  
	C_{(4-c)}^{[\bar{K}^\ast]}\mathcal{U}_{\omega}^{1/2}&=-\sqrt{\frac{1}{m_{\Xi_c}m_{\Lambda_c}}}\beta_B\lambda g_V^2\tilde{g}_{\psi s}\gamma_\omega\gamma_5 u_{\psi s},&
	C_{(4-c)}^{[\bar{K}^\ast]}\mathcal{U}_{\omega}^{3/2}&=-\sqrt{\frac{1}{m_{\Xi_c}m_{\Lambda_c}}}\beta_B\lambda g_V^2\tilde{g}_{\psi s}^{\prime}u_{\psi s  \omega},\nonumber\\ 
	C_{(4-d)}^{[\bar{K}^\ast]}\mathcal{U}_{\omega}^{1/2}&=-\frac{1}{4}\sqrt{\frac{1}{m_{\Xi_c}m_{\Lambda_c}}}\beta_Bg_V^2\tilde{g}_{\psi s}\gamma_\omega\gamma_5 u_{\psi s},&
	C_{(4-d)}^{[\bar{K}^\ast]}\mathcal{U}_{\omega}^{3/2}&=-\frac{1}{4}\sqrt{\frac{1}{m_{\Xi_c}m_{\Lambda_c}}}\beta_B\tilde{g}_{\psi s}^{\prime}u_{\psi s  \omega},\nonumber\\ 
	C_{(4-e)}^{[D]}\mathcal{U}_{\omega}^{1/2}&=-2\sqrt{\frac{2m_D^2m_{D^\ast}}{3m_{J/\psi}}}g_{B_8 B_{\bar{3}}D}g_2\tilde{g}_{\psi s}\gamma_\omega\gamma_5 u_{\psi s},&
	C_{(4-e)}^{[D]}\mathcal{U}_{\omega}^{3/2}&=-2\sqrt{\frac{2m_D^2m_{D^\ast}}{3m_{J/\psi}}}g_{B_8 B_{\bar{3}}D}g_2\tilde{g}_{\psi s}^{\prime}u_{\psi s  \omega},\nonumber\\ 
	C_{(4-e)}^{[D^\ast]}\mathcal{U}_{\omega}^{1/2}&=-\sqrt{\frac{2m_{D^\ast}^3m_{J/\psi}}{3}}g_{B_8 B_{\bar{3}}D^\ast}g_2\tilde{g}_{\psi s}\gamma_\omega\gamma_5 u_{\psi s},&
	C_{(4-e)}^{[D^\ast]}\mathcal{U}_{\omega}^{3/2}&=-\sqrt{\frac{2m_{D^\ast}^3m_{J/\psi}}{3}}g_{B_8 B_{\bar{3}}D^\ast}g_2\tilde{g}_{\psi s}^{\prime}u_{\psi s  \omega},\nonumber\\ 
	C_{(4-f)}^{[D]}\mathcal{U}_{\omega}^{1/2}&=\sqrt{\frac{2m_{D^\ast}m_D^2m_{\eta_c}}{3}}g_{B_8 B_{\bar{3}}D}g_2\tilde{g}_{\psi s}\gamma_\omega\gamma_5 u_{\psi s},&
	C_{(4-f)}^{[D]}\mathcal{U}_{\omega}^{3/2}&=\sqrt{\frac{2m_{D^\ast}m_D^2m_{\eta_c}}{3}}g_{B_8 B_{\bar{3}}D}g_2\tilde{g}_{\psi s}^{\prime}u_{\psi s  \omega},\nonumber\\ 
	C_{(4-f)}^{[D^\ast]}\mathcal{U}_{\omega}^{1/2}&=2\sqrt{\frac{2m_{D^\ast}^2m_Dm_{\eta_c}}{3}}g_{B_8 B_{\bar{3}}D^\ast}g_2\tilde{g}_{\psi s}\gamma_\omega\gamma_5 u_{\psi s},&
	C_{(4-f)}^{[D^\ast]}\mathcal{U}_{\omega}^{3/2}&=2\sqrt{\frac{2m_{D^\ast}^2m_Dm_{\eta_c}}{3}}g_{B_8 B_{\bar{3}}D^\ast}g_2\tilde{g}_{\psi s}^{\prime}u_{\psi s  \omega}.\nonumber\\ 
\end{align}	
Here, \(u_{\psi \omega}\) and \(u_{\psi s \omega}\) represent the spinors for the \([\Sigma_c \bar{D}^*]_{3/2}\) and \([\Xi_c \bar{D}^*]_{3/2}\) pentaquarks, respectively.
To facilitate practical application, the rather lengthy expressions for certain amplitudes have been simplified. The final forms are derived by performing the sum over the polarization vectors in the following manner:
\begin{eqnarray}
	 	\sum_{\lambda_1=-1,0,1}\varepsilon_{1\mu}^\dagger(q,\lambda_1)\varepsilon_{1\nu}(q,\lambda_1)&=&-g_{\mu\nu}+\frac{q_\mu q_\nu}{m_{\rho}^2},\\
	 	\sum_{\lambda_2=-1,0,1}\varepsilon_{2\mu}^\dagger(q,\lambda_2)\varepsilon_{2\nu}(q,\lambda_2)&=&-g_{\mu\nu}+\frac{q_\mu q_\nu}{m_{\bar{D}^{\ast}}^2},\\
	 	\sum_{\lambda_3=-1,0,1}\varepsilon_{3\mu}^\dagger(p_2-q,\lambda_3)\varepsilon_{3\nu}(p_2-q,\lambda_3)&=&-g_{\mu\nu}+\frac{(p_2-q)_\mu (p_2-q)_\nu}{m_{D^{\ast}}^2}.\\
	 	\sum_{\lambda_4=-1,0,1}\varepsilon_{4\mu}^\dagger(q,\lambda_4)\varepsilon_{4\nu}(q,\lambda_4)&=&-g_{\mu\nu}+\frac{q_\mu q_\nu}{m_{K^{\ast}}^2}.
\end{eqnarray}
	 
\end{widetext}
\end{appendices}

\bibliography{refs}
\end{document}